\documentclass{aastex701}

\usepackage{color}
\usepackage{amssymb}
\usepackage{graphicx}
\newcommand{\hpe}{Hz\,erg$^{-1}$}  %Hz per erg

\makeatletter
\g@addto@macro\@floatboxreset\centering

\shorttitle{}
\shortauthors{Hao et al.}

\makeatother

\begin{document}

\title{Gamma-ray bursts reveal the history and faint contributors of cosmic reionization}

\correspondingauthor{Jing-Meng Hao}

\author[orcid=0000-0001-8624-5989,gname=Jing-Meng,sname=Hao]{Jing-Meng Hao}
\affiliation{Department of Astronomy, School of Physics and Materials Science, Guangzhou University, Guangzhou 510006, China}
\affiliation{Dipartimento di Fisica e Astronomia, Universit\`{a} di Padova, Vicolo dell'Osservatorio, 3, I-35122 Padova, Italy}
\email[show]{jmhao@gzhu.edu.cn}

\author{Paolo Cassata}
\affiliation{Dipartimento di Fisica e Astronomia, Universit\`{a} di Padova, Vicolo dell'Osservatorio, 3, I-35122 Padova, Italy}
\affiliation{INAF –Osservatorio Astronomico di Padova, Vicolo dell’Osservatorio 5, I-35122 Padova, Italy}
\email{paolo.cassata@unipd.it}

\author{Zhen-Ya Zheng}
\affiliation{Shanghai Astronomical Observatory, CAS, Shanghai 200030, China}
\email{zhengzy@shao.ac.cn}

\author{Andrea Grazian}
\affiliation{INAF –Osservatorio Astronomico di Padova, Vicolo dell’Osservatorio 5, I-35122 Padova, Italy}
\email{andrea.grazian@inaf.it}

\author{Giulia Rodighiero}
\affiliation{Dipartimento di Fisica e Astronomia, Universit\`{a} di Padova, Vicolo dell'Osservatorio, 3, I-35122 Padova, Italy}
\affiliation{INAF –Osservatorio Astronomico di Padova, Vicolo dell’Osservatorio 5, I-35122 Padova, Italy}
\email{giulia.rodighiero@unipd.it}

\author{Alvio Renzini}
\affiliation{INAF –Osservatorio Astronomico di Padova, Vicolo dell’Osservatorio 5, I-35122 Padova, Italy}
\email{alvio.renzini@oapd.inaf.it}

\author{Jun-Hui Fan}
\affiliation{Department of Astronomy, School of Physics and Materials Science, Guangzhou University, Guangzhou 510006, China}
\email{fjh@gzhu.edu.cn}

\author{Andrea Ferrara}
\affiliation{Scuola Normale Superiore, Piazza dei Cavalieri 7, 56126 Pisa, Italy}
\email{andrea.ferrara@sns.it}

\begin{abstract}
Star-forming galaxies are generally believed to be the main drivers of
cosmic reionization.
However, the relative contributions of bright and faint galaxies
to this process remain unclear. As the most luminous transient phenomena
in the universe, long gamma-ray bursts (LGRBs) provide a unique
opportunity to probe star formation occurring in both detectable and undetectable
galaxies. In this Letter, we present new estimates of the cosmic star formation
rate density (SFRD) at $4<z<10$ using Swift LGRBs detected over the
past two decades, by considering LGRBs as unbiased tracers of total star formation at
high redshifts. Crucially, we find that the new LGRB-inferred SFRD can naturally explain current
measurements of hydrogen reionization without invoking extreme ionizing photon
production efficiencies or escape fractions from galaxies. Using these
LGRB-inferred SFRD values, we further investigate the faintest magnitude limits of
high-redshift galaxies, finding a redshift evolution of the limiting magnitudes
from $M_{\mathrm{lim}}\sim-14$ to $-15$ at $z\sim6$ to $M_{\mathrm{lim}}\sim-10$ to $-11$
at $z\sim10$. This result provides one independent piece of
evidence for the presence of a large population of faint galaxies at
redshifts $z\gtrsim6$, as an important complement to our understanding of
the ionizing photon budget in the early universe.
\end{abstract}


\section{Introduction\label{sec:Introduction}}

Over the past decade, measurements of the cosmic reionization history have advanced dramatically.
The electron scattering optical depth to the cosmic microwave background (CMB) from the
Planck satellite indicates this process was approximately halfway complete by $z\sim7-8$
\citep{2020A&A...641A...6P}. Lyman-$\alpha$ absorption constraints derived
from the spectra of distant quasars imply that the universe
was fully reionized at $5<z<6$ \citep{2022MNRAS.514...55B,2024MNRAS.533L..49Z,2026MNRAS.545f1862D}.
Given the minor contribution of quasars to the ionizing photons
\citep[e.g.][]{2022NatAs...6..850J,2025NatAs.tmp..200J}, it is generally believed
that hydrogen reionization is mainly driven by star-forming galaxies. However,
the relative contributions of bright and faint galaxies to the ionizing
radiation remain uncertain.

Recent James Webb Space Telescope (JWST) has extended the discovery of bright
galaxies out to redshifts as high as $z\sim14$, unexpectedly identifying an enhanced
abundance of UV bright galaxies \citep{2023Natur.622..707A,2024ApJ...969L...2F},
which seems to have unraveled parts of the mystery about the ionizing sources.
For example, some studies have suggested that these new galaxy observations, combined
with high ionizing efficiencies or large escape fractions, could infer sufficient, or even an
excess of ionizing photons. \citep[e.g.][]{2024MNRAS.535L..37M,2024MNRAS.535.2998S}.
While much work has focused on this bright galaxy population, the
numerous faint galaxies ($M_\mathrm{UV}\gtrsim-17$) which might dominate
star formation at redshifts $z\gtrsim6$ cannot be ignored. However, direct constraints on
fainter galaxies are still challenging even with the deepest JWST
surveys \citep{2023Natur.618..480R,2023ApJ...949L..25F,2024A&A...690A.288B}.
Although a number of studies have reported the discoveries of ultra-faint galaxies
through gravitational lensing \citep[e.g.][]{2019MNRAS.486.3805B,2024Natur.626..975A},
a complete understanding of how far the galaxy population extends toward the faint end is
still limited by small number statistics and uncertainties in the lensing modelling.

As the most luminous transient events in the observed universe, gamma-ray bursts (GRBs)
provide a unique and independent probe of high-redshift galaxies.
In particular, the physical link between long-duration GRBs (LGRBs) and short-lived
massive stars suggests that LGRBs can serve as tracers of total star formation
at high redshifts, potentially probing fainter galaxies than those accessible through direct
galaxy observations alone \citep[e.g.][]{2012ApJ...754...46T,2012ApJ...749L..38T}.
Indeed, analyses based on the LGRB redshift distribution have consistently indicated
a relatively high cosmic star formation rate density (SFRD) compared with estimates from galaxy surveys
\citep{2008ApJ...683L...5Y,2009ApJ...705L.104K,2011MNRAS.418..500I,2012ApJ...744...95R,2013arXiv1305.1630K}.
Still, the connection between the LGRB rate $\dot{n}_{\mathrm{LGRB}}(z)$ and the cosmic SFRD $\dot{\rho}_{\star}(z)$
needs to be treated with caution. In this respect, early studies \citep[e.g.][]{2009ApJ...705L.104K}
often parameterized the ratio of $\dot{n}_{\mathrm{LGRB}}(z)$ to $\dot{\rho}_{\star}(z)$ as a simple power law,
$\varepsilon(z)\propto(1+z)^{\alpha}$, with $\alpha\gtrsim1$ inferred from data at $z<4$.
This additional evolution of $\varepsilon(z)$ could be explained by the preference of LGRB progenitors
for low-metallicity environments, as predicted by the collapsar model \citep{2006ARA&A..44..507W}.
While the exact metallicity threshold for suppressing the LGRB production efficiency
remains debated, ranging from $0.3Z_{\bigodot}$ to near-solar value
\citep[e.g.][]{2013ApJ...772...42H,2016ApJ...817....8P,2020MNRAS.495..266M,2025A&A...703A.288D},
the situation is expected to be much simpler at high redshifts.

Since the mean metallicity of star-forming galaxies decreases rapidly with increasing
redshift up to $z\approx3$ \citep{2008A&A...488..463M}, even if a low-metallicity dependence
for LGRB production is present, the resulting enhancement in the LGRB rate derived
at low redshifts should be less important at higher redshifts. Indeed, in our previous work
we found that LGRBs appear to be unbiased tracers of the cosmic SFRD at redshifts
$z\gtrsim3$ \citep{2020ApJS..248...21H}. Moreover, observational studies
comparing the luminosity functions (LFs) of LGRB host galaxies with Lyman
break galaxies over $3<z<5$ \citep{2015ApJ...809...76G} and later
at $z\sim5$ \citep{2024ApJ...966..133S} also support the view that LGRBs
provide an unbiased estimator of star formation in all galaxies at high redshifts.

In this Letter, considering LGRBs as unbiased tracers of the cosmic
star formation history at $z\gtrsim3$, we attempt to present a robust
estimate of the SFRD at $z\simeq4-10$. Then, with this updated SFRD,
a further constraint on the faint-end magnitude limits of the galaxy LFs at
different redshifts could be executed. The Letter is organized as
follows. In Section \ref{sec:2}, we present the calculation of the
SFRD using Swift LGRBs. We then discuss the implications of our estimates
for reionization in Section \ref{sec:3}, and present our constraints
on the faintest limits of the galaxy LFs in Section \ref{sec:4}.
Finally, we give a brief summary of our work in Section \ref{sec:5}.
Throughout this Letter, we use the cosmological parameters from \citet{2020A&A...641A...6P}:
$\Omega_{\mathrm{m}}=0.315$, $\Omega_{\Lambda}=0.685$, $\Omega_{\mathrm{b}}=0.049$,
and $h=0.674$.

\section{Cosmic Star Formation Rate at High-redshifts\label{sec:2}}

In order to estimate the SFRD at $z\simeq4-10$ from high-redshift
LGRBs, we follow a similar method as proposed in \citet{2008ApJ...683L...5Y}.
Assuming the LGRB rate as an unbiased SFRD tracer at $z\gtrsim3$,
the expected redshift distribution of LGRBs can be related to
the SFRD as $\mathrm{d}N/\mathrm{d}z=A\dot{\rho}_{\star}(z)(\mathrm{d}V/\mathrm{d}z)/(1+z)$,
where the factor $(1+z)^{-1}$ accounts for the cosmological time dilation and
$\mathrm{d}V/\mathrm{d}z$ is the comoving volume per unit redshift. The constant $A$
is a factor that depends on the efficiency of the LGRB production and search
(e.g., the observing time, sky coverage, beaming factor, the detection limit of
LGRB luminosity, and the number of LGRB production per stellar mass). We calibrate
the value of $A$ using the observed number of LGRBs $N_{[3.5,\,4.5]}^\mathrm{obs}$
and the average value of SFRD measurements
$\left\langle \dot{\rho}_\star\right\rangle _{[3.5,\,4.5]}^\mathrm{obs}$
within the redshift range $z=3.5-4.5$. We then estimate the SFRD
$\left\langle \dot{\rho}_{\star}\right\rangle _{[z_{1},\,z_{2}]}$
within a higher redshift bin $z_{1}\leqslant z<z_{2}$ as
\begin{equation}
    \left\langle \dot{\rho}_{\star}\right\rangle _{[z_{1},\,z_{2}]}=
    \left\langle \dot{\rho}_{\star}\right\rangle _{[3.5,\,4.5]}^{\mathrm{obs}}
    \frac{N_{[z_{1},\,z_{2}]}^{\mathrm{obs}}}{N_{[3.5,\,4.5]}^{\mathrm{obs}}}
    \frac{\int_{3.5}^{4.5}\frac{\mathrm{d}V/\mathrm{d}z}{1+z}\mathrm{d}z}
    {\int_{z_{1}}^{z_{2}}\frac{\mathrm{d}V/\mathrm{d}z}{1+z}\mathrm{d}z},
    \label{eq:sfr}
\end{equation}
where $N_{[z_{1},\,z_{2}]}^{\mathrm{obs}}$ is the observed number
of LGRBs within the redshift bin $z_{1}\leqslant z<z_{2}$.
We choose the small redshift range $3.5\leqslant z<4.5$ as our calibration because:
(i) the SFRD is well constrained at redshift $z\sim4$; (ii) the effect of
the low-metallicity bias in the LGRB population is minimized at $z\gtrsim3$.

For the observed LGRB counts, we collect LGRBs detected by Swift
Burst Alert Telescope (BAT) over the past two decades (from 26 January
2005 to 5 December 2025), yielding a total of 486 events with measured redshifts.
Among them, there are 76 LGRBs at $z\geqslant3.5$ as listed in Table~\ref{tab1},
where the following parameters are included: the name, the redshift $z$ and its reference,
the duration $T_{90}$, the fluence $S_{\gamma}$ in 15-150~keV, the photon index $\Gamma$,
and the isotropic luminosity $L_{\mathrm{iso}}$.
We note that the $T_{90}$, $S_{\gamma}$, and $\Gamma$ values are taken from the third Swift BAT GRB
catalogue\footnote{\url{http://swift.gsfc.nasa.gov/results/batgrbcat/}} \citep{2016ApJ...829....7L},
supplemented by the Swift GRB table\footnote{\url{http://swift.gsfc.nasa.gov/archive/grb_table/}}.
The intrinsic isotropic luminosity for each burst can be calculated by
$L_{\mathrm{iso}}=E_{\mathrm{iso}}/[T_{90}/(1+z)]$,
where $E_\mathrm{iso}$ is the intrinsic isotropic equivalent energy computed by
\begin{equation}
E_{\mathrm{iso}}=\frac{4\pi d_{L}^{2}(z)S_{\gamma}}{1+z}k(z),
\end{equation}
where $k(z)$ is the \emph{k}-correction for the fluence transferred from the observed detector
band to its rest-frame. In order to reduce the uncertainty due to Swift's narrow bandpass,
we follow the method proposed by \citet{2016ApJ...817....7P}, where only
the energy release in the 45-450 keV rest-frame band is calculated.
Specifically, they calculated $k(z)$ using a simple form $k(z)=[(1+z)/(1+2)]^{\Gamma-2}$
with the measured photon index $\Gamma$, which makes the \emph{k}-correction always small
and the uncertainty even smaller.
The result for the luminosity-redshift distribution of our sample is shown in Figure~\ref{fig:lisoz},
where the shaded region approximates the effective
luminosity threshold for Swift detection, which depends on redshift:
\begin{equation}
L_{\lim}(z)=4\pi d_{\mathrm{L}}^{2}(z)F_{\mathrm{lim}}k(z),
\end{equation}
with a flux limit of $\sim F_{\mathrm{lim}}=1.0\times10^{-8}~\mathrm{erg~s^{-1}~cm^{-2}}$
and the population median value of $\Gamma=1.5$ for the calculation of $k(z)$.
We should stress here that the detection of a burst and its redshift measurement are complicated,
especially for the high-redshift bursts with fainter afterglows, making the redshift completeness of LGRB sample
subject to numerous selection effects. Nonetheless, \citet{2016ApJ...817....7P} found that the redshift
distribution of their complete SHOALS sample is very similar to that of the overall Swift LGRBs (see also \citealt{2020ApJS..248...21H}).
Hence, it is reasonable to use all known-redshift Swift LGRBs as our sample to maximize LGRB number
statistics in this work.
As shown in Figure~\ref{fig:lisoz}, high-redshift LGRBs with $z\geqslant3.5$ are then separated into subsamples used for estimating the SFRD
from Equation~(\ref{eq:sfr}). Since low-redshift, low-luminosity LGRBs would not be detected at
higher redshift, bursts within each redshift bin are compared only to the benchmark
subsample above the same luminosity cut. The threshold value at the center of each
redshift bin is used as the lower luminosity cut of the corresponding bin.
This approach avoids the need for a detail modeling of the LGRB
luminosity function above the detection thresholds.

Our inferred high-redshift SFRD values are shown in Figure~\ref{fig:sfrd}.
For the redshift bins $z=4.5-5.5$, $5.5-6.5$, $6.5-7.5$, $7.5-8.5$, $8.5-10$, we obtain
$\log[\dot{\rho}_{\star}/(\mathrm{M_{\odot}~yr^{-1}~Mpc^{-3}})]=-1.09_{-0.10}^{+0.09}$,
$-1.55_{-0.21}^{+0.18}$, $-1.77_{-0.36}^{+0.28}$, $-1.43_{-0.28}^{+0.23}$,
and $-1.89_{-0.57}^{+0.40}$, respectively. The error bars represent 68\%
Poisson confidence intervals for the number of events in each redshift bin
\citep{1991ApJ...374..344K}. With our new estimates of $\dot{\rho}_{\star}(z)$,
a simple log-linear fit at $z\geqslant4$ is also provided:
$\log\dot{\rho}_{\star}=k(z-4)+\log\dot{\rho}_{\star}^{z\sim4}$,
where $\log\dot{\rho}_{\star}^{z\sim4}=-1.10$ is the average observed value at
$z\sim4$ and $k=-0.15\pm0.02$ is the evolutionary parameter. It should be stressed
that, although the number of high-redshift LGRBs is not large, our results are
still significant, because the presence of even one burst would imply a sufficiently
high level of star formation at that redshift range, as discussed
in \citet{2009ApJ...705L.104K}. Since any excess detections of LGRBs
at $z>6$ only lead to even higher star formation rates, our inferred SFRD values
should be considered as conservative estimates.

Also shown in Figure~\ref{fig:sfrd} are recent observational data from the UV,
infrared (IR)/(sub)millimeter, and radio observations, including the JWST observations. Note that
most of the UV data shown here come from integrating the galaxy UV LFs down to $M_{\mathrm{lim}}=-17$,
the approximate faint limit of the UV observations \citep[e.g.][]{2014ARA&A..52..415M,2020ApJ...902..112B}.
The only exception is the data from \cite{2019MNRAS.486.3805B} with $M_{\mathrm{lim}}=-13.5$.
We find that, while our new estimates are broadly consistent with the observational
data at $z\sim4-6$, a systematic discrepancy (especially compared with the UV data)
becomes more pronounced from redshift $z\sim6$ to higher redshifts.
A possible explanation for the difference is that most of the star formation,
and the corresponding integrated UV luminosity, at high redshifts arises
in faint, undetected galaxies with $M_{\mathrm{UV}}\gtrsim-17$.
Obviously, this difference has potentially important implications for the reionization history of the universe.
Hence, the observational constraints of the reionization history provide a powerful test
to determine whether the proposition is correct, which is explored in the next section.
Since then, this difference could be further explored through a comparison of our LGRB-inferred
$\dot{\rho}_{\star}(z)$ and those derived from observed UV LFs of galaxies
assuming different limits (see Section~\ref{sec:4}).

\section{Implications for reionization\label{sec:3}}

To fully investigate our LGRB-derived SFRD with current measurements of the reionization history,
we calculate the evolution of the volume-filling fraction of ionized
hydrogen $Q_\mathrm{HII}$ and the optical depth of CMB photons
due to the Thomson scattering $\tau_{e}$. First, the value of $Q_{\mathrm{HII}}$ is
governed by the following differential equation:
\begin{equation}
    \frac{\mathrm{d}Q_{\mathrm{HII}}}{\mathrm{d}t}=\frac{\dot{n}_{\mathrm{ion}}}{\left\langle n_{\mathrm{H}}\right\rangle }-\frac{Q_{\mathrm{HII}}}{\bar{t}_{\mathrm{rec}}},
\end{equation}
where $\left\langle n_{\mathrm{H}}\right\rangle =1.88\times10^{-7}(\Omega_{\mathrm{b}}h^{2}/0.022)~\mathrm{cm^{-3}}$
is the mean hydrogen number density and the average recombination
time $\bar{t}_{\mathrm{rec}}$ is often represented as
\begin{equation}
    \bar{t}_{\mathrm{rec}}=\frac{1}{C_{\mathrm{HII}}\alpha_{\mathrm{B}}(T)(1+Y_{\mathrm{p}}/4X_{\mathrm{p}})\left\langle n_{\mathrm{H}}\right\rangle (1+z)^{3}},
\end{equation}
where $\alpha_{\mathrm{B}}=2.6\times10^{-13}~\mathrm{cm^{3}~s^{-1}}$
is the case B recombination coefficient of hydrogen at $T\simeq10^{4}\,\mathrm{T}$
\citep{2001PhR...349..125B}, $X_{\mathrm{p}}=0.75$ is the primordial mass fraction of
hydrogen, $Y_{\mathrm{p}}=1-X_{\mathrm{p}}$ is the primordial helium abundance.
Following the results of hydrodynamical simulations by \citet{2012ApJ...747..100S},
the clumping factor $C_{\mathrm{HII}}$ that accounts for inhomogeneities
in the intergalactic medium (IGM) is given by
\begin{equation}
    C_{\mathrm{HII}}=2.9\times\left[\frac{1+z}{6}\right]^{-1.1}.
\end{equation}

The ionizing emissivity $\dot{n}_\mathrm{ion}$ can be expressed as
$\dot{n}_\mathrm{ion}=f_\mathrm{esc}\xi_{\mathrm{ion}}\rho_{\mathrm{UV}}$,
where $\rho_{\mathrm{UV}}$ is the intrinsic UV luminosity density
of galaxies, $\xi_{\mathrm{ion}}$ is the ionizing photon production
efficiency representing the number of ionizing photons per unit UV
luminosity density, and $f_{\mathrm{esc}}$ is the escape fraction
of these ionizing photons from galaxies into the IGM. The intrinsic luminosity
density can be straightforwardly inferred from our LGRB-based SFRD (using the log-linear
fit and extrapolated to high redshifts) using the relation
\begin{equation}
L_{\mathrm{UV}}(\mathrm{erg\,s^{-1}\,Hz^{-1}})=
\mathrm{SFR}(\mathrm{M_{\odot}\,yr^{-1}})/\mathcal{K}_{\mathrm{UV}},
\label{eq:conv}
\end{equation}
with the conversion factor $\mathcal{K}_{\mathrm{UV}}=1.15\times10^{-28}$
as in \citet{2014ARA&A..52..415M}, assuming a Salpeter \citet{1955ApJ...121..161S} IMF.

The two parameters $\xi_\mathrm{ion}$ and $f_\mathrm{esc}$, however, remain elusive.
Many pre-JWST studies typically assumed a canonical value of
$\xi_\mathrm{ion}\simeq10^{25.2}$~\hpe in reionization
calculations \citep[e.g.][]{2015ApJ...802L..19R}, but early JWST measurements
suggested a high value of $\xi_\mathrm{ion}\simeq10^{25.8}$~\hpe
for faint galaxies during the epoch of reionization \citep{2024Natur.626..975A}.
More recent, deeper JWST surveys have revised these estimates downward, converging to
the typical values of pre-JWST once more, now suggesting an average efficiency of
$\xi_\mathrm{ion}\simeq10^{25.3}$~\hpe \citep{2024MNRAS.535.2998S,2025MNRAS.537.3245B,2025ApJ...981..134P}.
Similarly, most reionization models required a somewhat large value of $f_\mathrm{esc}\simeq0.2$
to complete reionization by $z\simeq6$ \cite[e.g.][]{2015ApJ...802L..19R,2025MNRAS.539.2409D},
whereas direct measurements of $f_\mathrm{esc}$ at $3<z<3.5$ reported typical values
of $f_\mathrm{esc}\simeq0.05$ \citep{2021MNRAS.505.2447P}.
It is also interesting to note a rather low value of $f_\mathrm{esc}\approx0.015$
constrained from GRB observations over $1.6<z<6.7$ \citep{2019MNRAS.483.5380T}.
At higher redshifts, recent studies suggested
an increase in escape fraction with redshift from $f_\mathrm{esc}\simeq0.07$
at $z\simeq10$ to $f_\mathrm{esc}\simeq0.19$ at $z\simeq11.8$,
based on the relationship between $f_\mathrm{esc}$ and the UV continuum
slope of high-redshift galaxies \citep{2024MNRAS.531..997C}.
Given these latest observational constraints, we choose the
median value of $\xi_\mathrm{ion}=10^{25.3}$~\hpe and a moderate escape
fraction $f_\mathrm{esc}=0.1$ as our fiducial values.

The resulting history of hydrogen reionization expressed through the redshift evolution
of the neutral hydrogen fraction $x_\mathrm{HI}=1-Q_{\mathrm{HII}}$ is shown in Figure~\ref{fig:EoR}.
We find that our predicted history of $x_\mathrm{HI}$ achieving reionization
at $z\sim6$, is in good agreement with the observational data and the mid-point redshift of
reionization $\left\langle z\right\rangle \simeq7.7\pm0.7$ from the Planck
measurements \citep{2020A&A...641A...6P}. The ionizing emissivity
$\dot{n}_\mathrm{ion}$ also has good consistency with the observational
constraints from \citet{2013MNRAS.436.1023B}, \citet{2023MNRAS.525.4093G}, and \citet{2024ApJ...969...12R}.
Moreover, as seen in Figure~\ref{fig:EoR}, we show two predictions of
our LGRB-inferred SFRD, assuming $f_\mathrm{esc}=0.1$ and
adopting the redshift-dependent forms for $\xi_\mathrm{ion}$ by
\citet{2025MNRAS.537.3245B} and \citet{2025A&A...698A.302L}, respectively.
Notably, all our results are clearly consistent with the available observational
constraints including the requirement that hydrogen reionization is
not fully completed until $z\simeq5.3$ \citep{2022MNRAS.514...55B}.

Second, the value of $\tau_{e}$ is an integral over the reionization history
to redshift $z$, given by
\begin{equation}
    \tau_{e}(z)=\int_{0}^{z}\frac{c(1+z')^{2}}{H(z')}Q_{\mathrm{HII}}\sigma_{\mathrm{T}}\left\langle n_{\mathrm{H}}\right\rangle (1+\eta Y/4X)\mathrm{d}z',\label{eq:tau}
\end{equation}
where $\sigma_{\mathrm{T}}$ is the Thomson electron scattering cross-section,
and $\eta$ gives the ionization state of helium. We take $\eta=1$ at $z>4$ and
$\eta=2$ at $z\leqslant4$ following \citet{2012MNRAS.423..862K}. We integrate the
optical depth over the reionization history to get the value of $\tau_{e}$ at $z=20$,
which is also fully consistent with the latest measurement
by \citet{2020A&A...641A...6P}, i.e.~$\tau_{e}=0.0544\pm0.0073$,
obtained from the CMB power spectra and lensing reconstruction (Figure~\ref{fig:tau}).

Our predictions on the timeline of hydrogen reionization and the Thomson scattering
optical depth suggest that hydrogen reionization can be naturally explained by our
LGRB-inferred SFRD fit without any need for dramatic assumptions on $\xi_\mathrm{ion}$
or $f_\mathrm{esc}$. Overall, these findings demonstrate the robustness and reliability
of our LGRBs-based SFRDs, implying an excess of star formation
occurring in faint galaxies at $z\gtrsim6$.

\section{Constraints on the faintest galaxies\label{sec:4}}

In order to analyze to what extent faint galaxies are below the detection limit at high redshifts,
we compare our LGRB-inferred SFRD fit of $\dot{\rho}_{\star}$ with the
conventional estimates determined from galaxy observations.
The conventional SFRD methods are typically derived from the UV LF of galaxies,
which is often parameterized as the Schechter function \citep{1976ApJ...203..297S}
with three parameters $\phi_{*}$, $M_{*}$, $\alpha$:
\begin{equation}
\phi(M_{\mathrm{UV}})=0.4\ln10\phi_{*}10^{0.4(\alpha+1)(M_{*}-M_{\mathrm{UV}})}\exp\left[-10^{0.4(M_{*}-M_{\mathrm{UV}})}\right].
\end{equation}
However, at $z\gtrsim6$, some authors reported an excess of bright galaxies relative
to the predictions by the classical Schechter function fits, suggesting
that the high-redshift LF may be better represented by a double power
law (DPL) shape with four parameters $\phi_{*}$, $M_{*}$, $\alpha$,
$\beta$ \citep[e.g.][]{2015MNRAS.452.1817B,2020MNRAS.493.2059B,2023MNRAS.518.6011D,2024MNRAS.533.3222D}:
\begin{equation}
\phi(M)=0.4\ln10\phi_{*}\left[10^{0.4(\alpha+1)(M_{\mathrm{UV}}-M_{*})}+10^{0.4(\beta+1)(M_{\mathrm{UV}}-M_{*})}\right]^{-1}.
\end{equation}
Then, the intrinsic UV luminosity density of galaxies $\rho_{\mathrm{UV}}$ brighter than the limiting magnitude $M_{\mathrm{lim}}$
can be calculated as follows:
\begin{equation}
    \rho_{\mathrm{UV}}=\int_{-\infty}^{M_{\mathrm{lim}}}\phi(M)L(M)10^{0.4\left\langle A_{\mathrm{UV}}\right\rangle }dM,
\end{equation}
where $\left\langle A_{\mathrm{UV}}\right\rangle $ is the average
dust attenuation given by
\begin{equation}
    \left\langle A_{\mathrm{UV}}\right\rangle =4.43+0.79\ln(10)\sigma_{\beta}^{2}+1.99\left\langle \beta\right\rangle ,
\end{equation}
with $\left\langle \beta\right\rangle $ being the UV-continuum spectrum
slope with a Gaussian dispersion $\sigma_{\beta}=0.34$ \citep{1999ApJ...521...64M,2013ApJ...768L..37T}. We estimate
$\beta$ following the linear $\beta-M_{\mathrm{UV}}$ relation from
\citet{2014ApJ...793..115B} for $z<8$ and adopt the most recent results
of \citet{2024MNRAS.531..997C} obtained from JWST for $z\geqslant8$.
Then we convert the luminosity density $\rho_{\mathrm{UV}}$
to the SFRD $\dot{\rho}_{\star}$ using Equation~(\ref{eq:conv}).

With a steep faint-end slope ($\alpha \gtrsim2$) for the observed UV LF at high
redshifts \citep[e.g.][]{2021AJ....162...47B}, the value of $\rho_\mathrm{UV}$ depends sensitively
on the faintest limiting magnitude of $M_\mathrm{lim}$. Instead of adopting
the commonly used, redshift-independent limit (e.g.~$M_{\mathrm{lim}}=-17$), we extend
the integral of LF to progressively fainter limits to match our LGRB-inferred SFRD fit.
The dust-corrected $\dot{\rho}_{\star}(z)$ values are obtained at redshifts
$\left\langle z\right\rangle =3.8,\,4.9,\,5.9,\,6.8,\,7.9,\,8.9,\,10.2$
using the best-fit Schechter functions from \citet{2021AJ....162...47B}, as shown in Figure~\ref{fig:sfrd}.
We find a gradual increase for limiting magnitudes from $M_\mathrm{lim}\sim-14$ at $z\sim6$ to
$M_\mathrm{lim}\sim-10$ at $z\sim10$.
For comparison, we also calculate the SFRD $\dot{\rho}_\star(z)$ implied by the DPL
fits from \citet{2022ApJS..259...20H} at $z\sim6-7$ and from \citet{2023MNRAS.518.6011D}
at $z\sim8-10$. This DPL model yields somewhat brighter limiting magnitudes, evolving
from $M_\mathrm{lim}\sim-15$ at $z\sim6$ to $M_\mathrm{lim}\sim-11$ at $z\sim10$,
due to the excess of bright galaxies. Anyway,
both models imply an evolution trend of the faint-end limiting magnitudes
of the LFs from redshift $z\sim6$ to $z\sim10$. We also note that, at $z\lesssim6$, the difference
of the LF-integrated SFRDs between the two limits $M_\mathrm{lim}=-17$ and $M_\mathrm{lim}=-10$ becomes smaller,
reflecting that faint galaxies have smaller contributions to the overall
star formation. These behaviors are consistent with a scenario in which star
formation in low-mass halos becomes increasingly suppressed by the feedback from
the UV background at lower redshifts \citep{2019MNRAS.488..419W,2021MNRAS.507.6108O}.

To further illustrate how much star formation occurs in galaxies above
the detection limit of current observations ($M_\mathrm{UV}\leqslant-17$),
we show in Figure~\ref{fig:fraction} the observable fraction of star formation,
$f_{\mathrm{obs}}(z)=\dot{\rho}_{\star}^{\mathrm{obs}}(z)/\dot{\rho}_{\star}^{\mathrm{tot}}(z)$,
defined as the ratio of the observable to the total SFRD, as a function of redshift
for both the Schechter function and DPL models. We find that $f_{\mathrm{obs}}(z)$
increases rapidly from $<30\%$ at $z\sim10$ to $>90\%$ after the completion
of reionization. This trend reflects a decline of the galaxy number density
at the faint end or a decreasing star formation efficiency in low-mass halos
at lower redshifts, as a result of the interplay between galaxy formation and
cosmic reionization. This result indicates that a large fraction of star formation
occurs in galaxies fainter than $M_\mathrm{UV}=-17$ at $z\gtrsim6$, contributing more
than $50\%$ of the total star formation at $z>8$.

\section{Summary\label{sec:5}}

In this Letter, by considering LGRBs as unbiased tracers of the total
star formation history at $z\gtrsim3$, we provide new estimates of the SFRD
$\dot{\rho}_{\star}(z)$ at $4<z<10$ using LGRB and SFRD data at $3.5<z<4.5$ as calibration.
These new $\dot{\rho}_{\star}(z)$ values exhibit a more gradual decline from
$z\sim4$ to $z\sim10$ compared to those determined from the visible galaxy population.
In particular, we find that they agree well with the observational
data at $z\sim4-6$, but show an excess at higher redshifts.
It is worth noting that, compared to previous LGRB studies \citep[e.g.][]{2013arXiv1305.1630K},
our results show qualitatively similar trends, but are systematically lower by $\sim0.1-0.5$ dex.
Combining with the latest JWST measurements of $\xi_\mathrm{ion}$ and $f_{\mathrm{esc}}$,
we find that our predictions are in remarkable agreement with current available
measurements of the cosmic reionization history, including the timeline of
hydrogen reionization and the Thomson scattering optical depth.

Our predictions of the excess of star formation at $z>6$
could be naturally explained by the presence of a large population of galaxies
being undetected at high redshifts. Hence, we take an additional
step of constraining the faint-end limiting magnitudes of the
galaxy LFs from our LGRB-inferred SFRD. When considering the UV LFs implied by a best-fit
Schechter model, we find a redshift evolution of the limiting magnitudes from
$M_{\mathrm{lim}}\sim-14$ at $z\sim6$ to $M_{\mathrm{lim}}\sim-10$ at $z\sim10$.
While for a DPL model, we find somewhat brighter values for the magnitude limits,
such as $M_{\mathrm{lim}}\sim-11$ at $z\sim10$. These imply that faint galaxies
($M_\mathrm{UV}>-17$) account for more than $50\%$ of the total star formation at $z>8$.
We should note that there are also other possible explanations. For example, the UV
estimates might have seriously underestimated the contribution of dust-obscured star formation,
since the IR estimates are systematically higher than UV estimates at $3\lesssim z\lesssim6$.
However, the exploration of this issue is beyond the scope of this Letter.
Finally, we stress that the number of LGRBs at high redshifts is still too small to make
any conclusive statement on the high-redshift SFRD. Fortunately, this situation
would soon be greatly alleviated, with the help of ``next-generation'' space missions such as
the Einstein Probe (EP;~\citealt{2025SCPMA..6839501Y}), the Space-based multi-band
astronomical Variable Objects Monitor (SVOM;~\citealt{2016arXiv161006892W}),
and the Transient High Energy Sky and Early Universe Surveyor
(THESEUS;~\citealt{2018AdSpR..62..191A}). For example, the THESEUS mission
is expected to detect more than 10 GRBs at $z>6$ every year \citep{2021ExA....52..219T}.
We also expect that, in the future, deep searches
for GRB hosts at high-redshifts could provide us a direct quantification
of the ratio of star formation in observed versus unobserved galaxies,
which would lead us to a better understanding of the nature of star
formation, and the properties of faint galaxies in the early universe.

\begin{acknowledgments}
We thank the anonymous referee for the detailed review and many helpful suggestions.
We thank Laura Bisigello for helpful discussions.
We acknowledge the use of public data from the Swift data archive.
J.-M.H. acknowledges support from the National Natural Science Foundation of China
(NSFC) (grant no. 11503004), and the China Scholarship Council program.
P.C., G.R., and A.G. are supported by the European
Union - NextGenerationEU RFF M4C2 1.1 PRIN 2022 project 2022ZSL4BL INSIGHT.
Z.-Y.Z. acknowledges the support by the China Manned Space Program (CMS-CSST-2025-A18).
A.G. acknowledges the support of the INAF GO/GTO Grant
2023 “Finding the Brightest Cosmic Beacons in the Universe with QUBRICS” (PI: Grazian).
J.-H.F. acknowledges support from the NSFC (grant no. 12433004), the Eighteenth Regular Meeting Exchange Project of
The Scientific and Technological Cooperation Committee between the People’s Republic of China and the
Republic of Bulgaria (series no. 1802) and National Key Research and Development Program of China (grant no. 2025YFA1614102).
A.F. is supported by the ERC Advanced Grant INTERSTELLAR H2020/740120, and in part by grant
NSF PHY-2309135 to the Kavli Institute for Theoretical Physics.
\end{acknowledgments}

\bibliographystyle{aasjournalv7}
\bibliography{refgrb}

@ARTICLE{2015ApJ...804...51C,
  author = {{Cucchiara}, A. and {Fumagalli}, M. and {Rafelski}, M. and {Kocevski},
	D. and {Prochaska}, J.~X. and {Cooke}, R.~J. and {Becker}, G.~D.},
  title = {Unveiling the Secrets of Metallicity and Massive Star Formation Using
	DLAs along Gamma-Ray Bursts},
  journal = {\apj},
  year = {2015},
  volume = {804},
  pages = {51},
  month = may,
  eid = {51},
  adsurl = {http://adsabs.harvard.edu/abs/2015ApJ...804...51C},
  archiveprefix = {arXiv},
  doi = {10.1088/0004-637X/804/1/51},
  eprint = {1408.3578}
}

@ARTICLE{2011ApJ...736....7C,
  author = {{Cucchiara}, A. and {Levan}, A.~J. and {Fox}, D.~B. and {Tanvir},
	N.~R. and {Ukwatta}, T.~N. and {Berger}, E. and {Kr{\"u}hler}, T.
	and {K{\"u}pc{\"u} Yolda{\c s}}, A. and {Wu}, X.~F. and {Toma}, K.
	and {Greiner}, J. and {Olivares}, F.~E. and {Rowlinson}, A. and {Amati},
	L. and {Sakamoto}, T. and {Roth}, K. and {Stephens}, A. and {Fritz},
	A. and {Fynbo}, J.~P.~U. and {Hjorth}, J. and {Malesani}, D. and
	{Jakobsson}, P. and {Wiersema}, K. and {O'Brien}, P.~T. and {Soderberg},
	A.~M. and {Foley}, R.~J. and {Fruchter}, A.~S. and {Rhoads}, J. and
	{Rutledge}, R.~E. and {Schmidt}, B.~P. and {Dopita}, M.~A. and {Podsiadlowski},
	P. and {Willingale}, R. and {Wolf}, C. and {Kulkarni}, S.~R. and
	{D'Avanzo}, P.},
  title = {A Photometric Redshift of z \~{} 9.4 for GRB 090429B},
  journal = {\apj},
  year = {2011},
  volume = {736},
  pages = {7},
  month = jul,
  eid = {7},
  adsurl = {http://adsabs.harvard.edu/abs/2011ApJ...736....7C},
  archiveprefix = {arXiv},
  doi = {10.1088/0004-637X/736/1/7},
  eprint = {1105.4915}
}

@ARTICLE{2015ApJ...809...76G,
  author = {{Greiner}, J. and {Fox}, D.~B. and {Schady}, P. and {Kr{\"u}hler},
	T. and {Trenti}, M. and {Cikota}, A. and {Bolmer}, J. and {Elliott},
	J. and {Delvaux}, C. and {Perna}, R. and {Afonso}, P. and {Kann},
	D.~A. and {Klose}, S. and {Savaglio}, S. and {Schmidl}, S. and {Schweyer},
	T. and {Tanga}, M. and {Varela}, K.},
  title = {Gamma-Ray Bursts Trace UV Metrics of Star Formation over 3 $\lt$
	z $\lt$ 5},
  journal = {\apj},
  year = {2015},
  volume = {809},
  pages = {76},
  month = aug,
  eid = {76},
  adsurl = {http://adsabs.harvard.edu/abs/2015ApJ...809...76G},
  archiveprefix = {arXiv},
  doi = {10.1088/0004-637X/809/1/76},
  eprint = {1503.05323},
  primaryclass = {astro-ph.HE}
}

@ARTICLE{2020ApJS..248...21H,
  author = {{Hao}, Jing-Meng and {Cao}, Liang and {Lu}, You-Jun and {Chu}, Qing-Bo
	and {Fan}, Jun-Hui and {Yuan}, Ye-Fei and {Yuan}, Yu-Hai},
  title = {Revisiting the Relationship between the Long GRB Rate and Cosmic
	Star Formation History Based on a Large Swift Sample},
  journal = {\apjs},
  year = {2020},
  volume = {248},
  pages = {21},
  number = {1},
  month = may,
  eid = {21},
  adsurl = {https://ui.adsabs.harvard.edu/abs/2020ApJS..248...21H},
  archiveprefix = {arXiv},
  doi = {10.3847/1538-4365/ab88da},
  eprint = {2005.07630},
  primaryclass = {astro-ph.HE}
}

@ARTICLE{2013ApJ...772...42H,
  author = {{Hao}, J.-M. and {Yuan}, Y.-F.},
  title = {Is the Metallicity of the Progenitor of Long Gamma-Ray Bursts Really
	Low?},
  journal = {\apj},
  year = {2013},
  volume = {772},
  pages = {42},
  month = jul,
  eid = {42},
  adsurl = {http://adsabs.harvard.edu/abs/2013ApJ...772...42H},
  archiveprefix = {arXiv},
  doi = {10.1088/0004-637X/772/1/42},
  eprint = {1305.5165},
  primaryclass = {astro-ph.HE}
}

@ARTICLE{2009ApJ...705L.104K,
  author = {{Kistler}, M.~D. and {Y{\"u}ksel}, H. and {Beacom}, J.~F. and {Hopkins},
	A.~M. and {Wyithe}, J.~S.~B.},
  title = {The Star Formation Rate in the Reionization Era as Indicated by Gamma-Ray
	Bursts},
  journal = {\apjl},
  year = {2009},
  volume = {705},
  pages = {L104-L108},
  month = nov,
  adsurl = {http://adsabs.harvard.edu/abs/2009ApJ...705L.104K},
  archiveprefix = {arXiv},
  doi = {10.1088/0004-637X/705/2/L104},
  eprint = {0906.0590},
  primaryclass = {astro-ph.CO}
}

@Article{2016ApJ...829....7L,
  author        = {{Lien}, A. and {Sakamoto}, T. and {Barthelmy}, S.~D. and {Baumgartner}, W.~H. and {Cannizzo}, J.~K. and {Chen}, K. and {Collins}, N.~R. and {Cummings}, J.~R. and {Gehrels}, N. and {Krimm}, H.~A. and {Markwardt}, C.~B. and {Palmer}, D.~M. and {Stamatikos}, M. and {Troja}, E. and {Ukwatta}, T.~N.},
  journal       = {\apj},
  title         = {The Third Swift Burst Alert Telescope Gamma-Ray Burst Catalog},
  year          = {2016},
  month         = sep,
  pages         = {7},
  volume        = {829},
  adsurl        = {http://adsabs.harvard.edu/abs/2016ApJ...829....7L},
  archiveprefix = {arXiv},
  doi           = {10.3847/0004-637X/829/1/7},
  eid           = {7},
  eprint        = {1606.01956},
  primaryclass  = {astro-ph.HE},
}

@ARTICLE{2014ARA&A..52..415M,
  author = {{Madau}, P. and {Dickinson}, M.},
  title = {Cosmic Star-Formation History},
  journal = {\araa},
  year = {2014},
  volume = {52},
  pages = {415-486},
  month = aug,
  adsurl = {http://adsabs.harvard.edu/abs/2014ARA%26A..52..415M},
  archiveprefix = {arXiv},
  doi = {10.1146/annurev-astro-081811-125615},
  eprint = {1403.0007}
}

@ARTICLE{2017A&A...602A...5N,
  author = {{Novak}, M. and {Smol{\v{c}}i{\'c}}, V. and {Delhaize}, J. and {Delvecchio},
	I. and {Zamorani}, G. and {Baran}, N. and {Bondi}, M. and {Capak},
	P. and {Carilli}, C.~L. and {Ciliegi}, P. and {Civano}, F. and {Ilbert},
	O. and {Karim}, A. and {Laigle}, C. and {Le F{\`e}vre}, O. and {Marchesi},
	S. and {McCracken}, H. and {Miettinen}, O. and {Salvato}, M. and
	{Sargent}, M. and {Schinnerer}, E. and {Tasca}, L.},
  title = {The VLA-COSMOS 3 GHz Large Project: Cosmic star formation history
	since z 5},
  journal = {\aap},
  year = {2017},
  volume = {602},
  pages = {A5},
  month = Jun,
  eid = {A5},
  adsurl = {https://ui.adsabs.harvard.edu/#abs/2017A&A...602A...5N},
  doi = {10.1051/0004-6361/201629436},
  primaryclass = {Astrophysics - Astrophysics of Galaxies}
}

@ARTICLE{2016ApJ...817....7P,
  author = {{Perley}, D.~A. and {Kr{\"u}hler}, T. and {Schulze}, S. and {de Ugarte
	Postigo}, A. and {Hjorth}, J. and {Berger}, E. and {Cenko}, S.~B.
	and {Chary}, R. and {Cucchiara}, A. and {Ellis}, R. and {Fong}, W.
	and {Fynbo}, J.~P.~U. and {Gorosabel}, J. and {Greiner}, J. and {Jakobsson},
	P. and {Kim}, S. and {Laskar}, T. and {Levan}, A.~J. and {Micha{\l}owski},
	M.~J. and {Milvang-Jensen}, B. and {Tanvir}, N.~R. and {Th{\"o}ne},
	C.~C. and {Wiersema}, K.},
  title = {The Swift Gamma-Ray Burst Host Galaxy Legacy Survey. I. Sample Selection
	and Redshift Distribution},
  journal = {\apj},
  year = {2016},
  volume = {817},
  pages = {7},
  month = jan,
  eid = {7},
  adsurl = {http://adsabs.harvard.edu/abs/2016ApJ...817....7P},
  archiveprefix = {arXiv},
  doi = {10.3847/0004-637X/817/1/7},
  eprint = {1504.02482}
}

@ARTICLE{2016ApJ...817....8P,
  author = {{Perley}, D.~A. and {Tanvir}, N.~R. and {Hjorth}, J. and {Laskar},
	T. and {Berger}, E. and {Chary}, R. and {de Ugarte Postigo}, A. and
	{Fynbo}, J.~P.~U. and {Kr{\"u}hler}, T. and {Levan}, A.~J. and {Micha{\l}owski},
	M.~J. and {Schulze}, S.},
  title = {The Swift GRB Host Galaxy Legacy Survey. II. Rest-frame Near-IR Luminosity
	Distribution and Evidence for a Near-solar Metallicity Threshold},
  journal = {\apj},
  year = {2016},
  volume = {817},
  pages = {8},
  month = jan,
  eid = {8},
  adsurl = {http://adsabs.harvard.edu/abs/2016ApJ...817....8P},
  archiveprefix = {arXiv},
  doi = {10.3847/0004-637X/817/1/8},
  eprint = {1504.02479}
}

@Article{2012ApJ...744...95R,
  author        = {{Robertson}, B.~E. and {Ellis}, R.~S.},
  journal       = {\apj},
  title         = {Connecting the Gamma Ray Burst Rate and the Cosmic Star Formation History: Implications for Reionization and Galaxy Evolution},
  year          = {2012},
  month         = jan,
  pages         = {95},
  volume        = {744},
  archiveprefix = {arXiv},
  doi           = {10.1088/0004-637X/744/2/95},
  eid           = {95},
  eprint        = {1109.0990},
  primaryclass  = {astro-ph.CO},
  url           = {http://adsabs.harvard.edu/abs/2012ApJ...744...95R},
}

@ARTICLE{1955ApJ...121..161S,
  author = {{Salpeter}, E.~E.},
  title = {The Luminosity Function and Stellar Evolution.},
  journal = {\apj},
  year = {1955},
  volume = {121},
  pages = {161},
  month = jan,
  adsurl = {http://adsabs.harvard.edu/abs/1955ApJ...121..161S},
  doi = {10.1086/145971}
}

@ARTICLE{1976ApJ...203..297S,
  author = {{Schechter}, P.},
  title = {An analytic expression for the luminosity function for galaxies.},
  journal = {\apj},
  year = {1976},
  volume = {203},
  pages = {297-306},
  month = jan,
  adsurl = {https://ui.adsabs.harvard.edu/abs/1976ApJ...203..297S},
  doi = {10.1086/154079}
}

@ARTICLE{2009Natur.461.1254T,
  author = {{Tanvir}, N.~R. and {Fox}, D.~B. and {Levan}, A.~J. and {Berger},
	E. and {Wiersema}, K. and {Fynbo}, J.~P.~U. and {Cucchiara}, A. and
	{Kr{\"u}hler}, T. and {Gehrels}, N. and {Bloom}, J.~S. and {Greiner},
	J. and {Evans}, P.~A. and {Rol}, E. and {Olivares}, F. and {Hjorth},
	J. and {Jakobsson}, P. and {Farihi}, J. and {Willingale}, R. and
	{Starling}, R.~L.~C. and {Cenko}, S.~B. and {Perley}, D. and {Maund},
	J.~R. and {Duke}, J. and {Wijers}, R.~A.~M.~J. and {Adamson}, A.~J.
	and {Allan}, A. and {Bremer}, M.~N. and {Burrows}, D.~N. and {Castro-Tirado},
	A.~J. and {Cavanagh}, B. and {de Ugarte Postigo}, A. and {Dopita},
	M.~A. and {Fatkhullin}, T.~A. and {Fruchter}, A.~S. and {Foley},
	R.~J. and {Gorosabel}, J. and {Kennea}, J. and {Kerr}, T. and {Klose},
	S. and {Krimm}, H.~A. and {Komarova}, V.~N. and {Kulkarni}, S.~R.
	and {Moskvitin}, A.~S. and {Mundell}, C.~G. and {Naylor}, T. and
	{Page}, K. and {Penprase}, B.~E. and {Perri}, M. and {Podsiadlowski},
	P. and {Roth}, K. and {Rutledge}, R.~E. and {Sakamoto}, T. and {Schady},
	P. and {Schmidt}, B.~P. and {Soderberg}, A.~M. and {Sollerman}, J.
	and {Stephens}, A.~W. and {Stratta}, G. and {Ukwatta}, T.~N. and
	{Watson}, D. and {Westra}, E. and {Wold}, T. and {Wolf}, C.},
  title = {A {$\gamma$}-ray burst at a redshift of z\~{}8.2},
  journal = {\nat},
  year = {2009},
  volume = {461},
  pages = {1254-1257},
  month = oct,
  adsurl = {http://adsabs.harvard.edu/abs/2009Natur.461.1254T},
  archiveprefix = {arXiv},
  doi = {10.1038/nature08459},
  eprint = {0906.1577},
  primaryclass = {astro-ph.CO}
}

@ARTICLE{2012ApJ...749L..38T,
  author = {{Trenti}, M. and {Perna}, R. and {Levesque}, E.~M. and {Shull}, J.~M.
	and {Stocke}, J.~T.},
  title = {Gamma-Ray Burst Host Galaxy Surveys at Redshift z {\gt}\~{} 4: Probes
	of Star Formation Rate and Cosmic Reionization},
  journal = {\apjl},
  year = {2012},
  volume = {749},
  pages = {L38},
  month = apr,
  eid = {L38},
  adsurl = {http://adsabs.harvard.edu/abs/2012ApJ...749L..38T},
  archiveprefix = {arXiv},
  doi = {10.1088/2041-8205/749/2/L38},
  eprint = {1202.0010},
  primaryclass = {astro-ph.CO}
}

@ARTICLE{2006ARA&A..44..507W,
  author = {{Woosley}, S.~E. and {Bloom}, J.~S.},
  title = {The Supernova Gamma-Ray Burst Connection},
  journal = {\araa},
  year = {2006},
  volume = {44},
  pages = {507-556},
  month = sep,
  adsurl = {http://adsabs.harvard.edu/abs/2006ARA%26A..44..507W},
  doi = {10.1146/annurev.astro.43.072103.150558},
  eprint = {arXiv:astro-ph/0609142}
}

@ARTICLE{2008ApJ...683L...5Y,
  author = {{Y{\"u}ksel}, H. and {Kistler}, M.~D. and {Beacom}, J.~F. and {Hopkins},
	A.~M.},
  title = {Revealing the High-Redshift Star Formation Rate with Gamma-Ray Bursts},
  journal = {\apjl},
  year = {2008},
  volume = {683},
  pages = {L5-L8},
  month = aug,
  adsurl = {http://adsabs.harvard.edu/abs/2008ApJ...683L...5Y},
  archiveprefix = {arXiv},
  doi = {10.1086/591449},
  eprint = {0804.4008}
}

@Article{2022NatAs...6..850J,
  author        = {{Jiang}, Linhua and {Ning}, Yuanhang and {Fan}, Xiaohui and {Ho}, Luis C. and {Luo}, Bin and {Wang}, Feige and {Wu}, Jin and {Wu}, Xue-Bing and {Yang}, Jinyi and {Zheng}, Zhen-Ya},
  journal       = {Nature Astronomy},
  title         = {{Definitive upper bound on the negligible contribution of quasars to cosmic reionization}},
  year          = {2022},
  month         = jun,
  pages         = {850-856},
  volume        = {6},
  adsurl        = {https://ui.adsabs.harvard.edu/abs/2022NatAs...6..850J},
  archiveprefix = {arXiv},
  doi           = {10.1038/s41550-022-01708-w},
  eprint        = {2206.07825},
  primaryclass  = {astro-ph.GA},
}

@Article{2022MNRAS.514...55B,
  author        = {{Bosman}, Sarah E.~I. and {Davies}, Frederick B. and {Becker}, George D. and {Keating}, Laura C. and {Davies}, Rebecca L. and {Zhu}, Yongda and {Eilers}, Anna-Christina and {D'Odorico}, Valentina and {Bian}, Fuyan and {Bischetti}, Manuela and {Cristiani}, Stefano V. and {Fan}, Xiaohui and {Farina}, Emanuele P. and {Haehnelt}, Martin G. and {Hennawi}, Joseph F. and {Kulkarni}, Girish and {Mesinger}, Andrei and {Meyer}, Romain A. and {Onoue}, Masafusa and {Pallottini}, Andrea and {Qin}, Yuxiang and {Ryan-Weber}, Emma and {Schindler}, Jan-Torge and {Walter}, Fabian and {Wang}, Feige and {Yang}, Jinyi},
  journal       = {\mnras},
  title         = {{Hydrogen reionization ends by z = 5.3: Lyman-{\ensuremath{\alpha}} optical depth measured by the XQR-30 sample}},
  year          = {2022},
  month         = jul,
  number        = {1},
  pages         = {55-76},
  volume        = {514},
  adsurl        = {https://ui.adsabs.harvard.edu/abs/2022MNRAS.514...55B},
  archiveprefix = {arXiv},
  doi           = {10.1093/mnras/stac1046},
  eprint        = {2108.03699},
  primaryclass  = {astro-ph.CO},
}

@Article{2020A&A...641A...6P,
  author        = {{Planck Collaboration} and {Aghanim}, N. and {Akrami}, Y. and {Ashdown}, M. and {Aumont}, J. and {Baccigalupi}, C. and {Ballardini}, M. and {Banday}, A.~J. and {Barreiro}, R.~B. and {Bartolo}, N. and {Basak}, S. and {Battye}, R. and {Benabed}, K. and {Bernard}, J. -P. and {Bersanelli}, M. and {Bielewicz}, P. and {Bock}, J.~J. and {Bond}, J.~R. and {Borrill}, J. and {Bouchet}, F.~R. and {Boulanger}, F. and {Bucher}, M. and {Burigana}, C. and {Butler}, R.~C. and {Calabrese}, E. and {Cardoso}, J. -F. and {Carron}, J. and {Challinor}, A. and {Chiang}, H.~C. and {Chluba}, J. and {Colombo}, L.~P.~L. and {Combet}, C. and {Contreras}, D. and {Crill}, B.~P. and {Cuttaia}, F. and {de Bernardis}, P. and {de Zotti}, G. and {Delabrouille}, J. and {Delouis}, J. -M. and {Di Valentino}, E. and {Diego}, J.~M. and {Dor{\'e}}, O. and {Douspis}, M. and {Ducout}, A. and {Dupac}, X. and {Dusini}, S. and {Efstathiou}, G. and {Elsner}, F. and {En{\ss}lin}, T.~A. and {Eriksen}, H.~K. and {Fantaye}, Y. and {Farhang}, M. and {Fergusson}, J. and {Fernandez-Cobos}, R. and {Finelli}, F. and {Forastieri}, F. and {Frailis}, M. and {Fraisse}, A.~A. and {Franceschi}, E. and {Frolov}, A. and {Galeotta}, S. and {Galli}, S. and {Ganga}, K. and {G{\'e}nova-Santos}, R.~T. and {Gerbino}, M. and {Ghosh}, T. and {Gonz{\'a}lez-Nuevo}, J. and {G{\'o}rski}, K.~M. and {Gratton}, S. and {Gruppuso}, A. and {Gudmundsson}, J.~E. and {Hamann}, J. and {Handley}, W. and {Hansen}, F.~K. and {Herranz}, D. and {Hildebrandt}, S.~R. and {Hivon}, E. and {Huang}, Z. and {Jaffe}, A.~H. and {Jones}, W.~C. and {Karakci}, A. and {Keih{\"a}nen}, E. and {Keskitalo}, R. and {Kiiveri}, K. and {Kim}, J. and {Kisner}, T.~S. and {Knox}, L. and {Krachmalnicoff}, N. and {Kunz}, M. and {Kurki-Suonio}, H. and {Lagache}, G. and {Lamarre}, J. -M. and {Lasenby}, A. and {Lattanzi}, M. and {Lawrence}, C.~R. and {Le Jeune}, M. and {Lemos}, P. and {Lesgourgues}, J. and {Levrier}, F. and {Lewis}, A. and {Liguori}, M. and {Lilje}, P.~B. and {Lilley}, M. and {Lindholm}, V. and {L{\'o}pez-Caniego}, M. and {Lubin}, P.~M. and {Ma}, Y. -Z. and {Mac{\'\i}as-P{\'e}rez}, J.~F. and {Maggio}, G. and {Maino}, D. and {Mandolesi}, N. and {Mangilli}, A. and {Marcos-Caballero}, A. and {Maris}, M. and {Martin}, P.~G. and {Martinelli}, M. and {Mart{\'\i}nez-Gonz{\'a}lez}, E. and {Matarrese}, S. and {Mauri}, N. and {McEwen}, J.~D. and {Meinhold}, P.~R. and {Melchiorri}, A. and {Mennella}, A. and {Migliaccio}, M. and {Millea}, M. and {Mitra}, S. and {Miville-Desch{\^e}nes}, M. -A. and {Molinari}, D. and {Montier}, L. and {Morgante}, G. and {Moss}, A. and {Natoli}, P. and {N{\o}rgaard-Nielsen}, H.~U. and {Pagano}, L. and {Paoletti}, D. and {Partridge}, B. and {Patanchon}, G. and {Peiris}, H.~V. and {Perrotta}, F. and {Pettorino}, V. and {Piacentini}, F. and {Polastri}, L. and {Polenta}, G. and {Puget}, J. -L. and {Rachen}, J.~P. and {Reinecke}, M. and {Remazeilles}, M. and {Renzi}, A. and {Rocha}, G. and {Rosset}, C. and {Roudier}, G. and {Rubi{\~n}o-Mart{\'\i}n}, J.~A. and {Ruiz-Granados}, B. and {Salvati}, L. and {Sandri}, M. and {Savelainen}, M. and {Scott}, D. and {Shellard}, E.~P.~S. and {Sirignano}, C. and {Sirri}, G. and {Spencer}, L.~D. and {Sunyaev}, R. and {Suur-Uski}, A. -S. and {Tauber}, J.~A. and {Tavagnacco}, D. and {Tenti}, M. and {Toffolatti}, L. and {Tomasi}, M. and {Trombetti}, T. and {Valenziano}, L. and {Valiviita}, J. and {Van Tent}, B. and {Vibert}, L. and {Vielva}, P. and {Villa}, F. and {Vittorio}, N. and {Wandelt}, B.~D. and {Wehus}, I.~K. and {White}, M. and {White}, S.~D.~M. and {Zacchei}, A. and {Zonca}, A.},
  journal       = {\aap},
  title         = {{Planck 2018 results. VI. Cosmological parameters}},
  year          = {2020},
  month         = sep,
  pages         = {A6},
  volume        = {641},
  adsurl        = {https://ui.adsabs.harvard.edu/abs/2020A&A...641A...6P},
  archiveprefix = {arXiv},
  doi           = {10.1051/0004-6361/201833910},
  eid           = {A6},
  eprint        = {1807.06209},
  primaryclass  = {astro-ph.CO},
}

@Article{2023ApJ...949L..25F,
  author        = {{Fujimoto}, Seiji and {Arrabal Haro}, Pablo and {Dickinson}, Mark and {Finkelstein}, Steven L. and {Kartaltepe}, Jeyhan S. and {Larson}, Rebecca L. and {Burgarella}, Denis and {Bagley}, Micaela B. and {Behroozi}, Peter and {Chworowsky}, Katherine and {Hirschmann}, Michaela and {Trump}, Jonathan R. and {Wilkins}, Stephen M. and {Yung}, L.~Y. Aaron and {Koekemoer}, Anton M. and {Papovich}, Casey and {Pirzkal}, Nor and {Ferguson}, Henry C. and {Fontana}, Adriano and {Grogin}, Norman A. and {Grazian}, Andrea and {Kewley}, Lisa J. and {Kocevski}, Dale D. and {Lotz}, Jennifer M. and {Pentericci}, Laura and {Ravindranath}, Swara and {Somerville}, Rachel S. and {Wilkins}, Stephen M. and {Amor{\'\i}n}, Ricardo O. and {Backhaus}, Bren E. and {Calabr{\`o}}, Antonello and {Casey}, Caitlin M. and {Cooper}, M.~C. and {Fern{\'a}ndez}, Vital and {Franco}, Maximilien and {Giavalisco}, Mauro and {Hathi}, Nimish P. and {Harish}, Santosh and {Hutchison}, Taylor A. and {Iyer}, Kartheik G. and {Jung}, Intae and {Lucas}, Ray A. and {Zavala}, Jorge A.},
  journal       = {\apjl},
  title         = {{CEERS Spectroscopic Confirmation of NIRCam-selected z {\ensuremath{\gtrsim}} 8 Galaxy Candidates with JWST/NIRSpec: Initial Characterization of Their Properties}},
  year          = {2023},
  month         = jun,
  number        = {2},
  pages         = {L25},
  volume        = {949},
  adsurl        = {https://ui.adsabs.harvard.edu/abs/2023ApJ...949L..25F},
  archiveprefix = {arXiv},
  doi           = {10.3847/2041-8213/acd2d9},
  eid           = {L25},
  eprint        = {2301.09482},
  primaryclass  = {astro-ph.GA},
}

@Article{2024A&A...690A.288B,
  author        = {{Bunker}, Andrew J. and {Cameron}, Alex J. and {Curtis-Lake}, Emma and {Jakobsen}, Peter and {Carniani}, Stefano and {Curti}, Mirko and {Witstok}, Joris and {Maiolino}, Roberto and {D'Eugenio}, Francesco and {Looser}, Tobias J. and {Willott}, Chris and {Bonaventura}, Nina and {Hainline}, Kevin and {{\"U}bler}, Hannah and {Willmer}, Christopher N.~A. and {Saxena}, Aayush and {Smit}, Renske and {Alberts}, Stacey and {Arribas}, Santiago and {Baker}, William M. and {Baum}, Stefi and {Bhatawdekar}, Rachana and {Bowler}, Rebecca A.~A. and {Boyett}, Kristan and {Charlot}, Stephane and {Chen}, Zuyi and {Chevallard}, Jacopo and {Circosta}, Chiara and {DeCoursey}, Christa and {de Graaff}, Anna and {Egami}, Eiichi and {Eisenstein}, Daniel J. and {Endsley}, Ryan and {Ferruit}, Pierre and {Giardino}, Giovanna and {Hausen}, Ryan and {Helton}, Jakob M. and {Hviding}, Raphael E. and {Ji}, Zhiyuan and {Johnson}, Benjamin D. and {Jones}, Gareth C. and {Kumari}, Nimisha and {Laseter}, Isaac and {L{\"u}tzgendorf}, Nora and {Maseda}, Michael V. and {Nelson}, Erica and {Parlanti}, Eleonora and {Perna}, Michele and {Rauscher}, Bernard J. and {Rawle}, Tim and {Rix}, Hans-Walter and {Rieke}, Marcia and {Robertson}, Brant and {Rodr{\'\i}guez Del Pino}, Bruno and {Sandles}, Lester and {Scholtz}, Jan and {Sharpe}, Katherine and {Skarbinski}, Maya and {Stark}, Daniel P. and {Sun}, Fengwu and {Tacchella}, Sandro and {Topping}, Michael W. and {Villanueva}, Natalia C. and {Wallace}, Imaan E.~B. and {Williams}, Christina C. and {Woodrum}, Charity},
  journal       = {\aap},
  title         = {{JADES NIRSpec initial data release for the Hubble Ultra Deep Field: Redshifts and line fluxes of distant galaxies from the deepest JWST Cycle 1 NIRSpec multi-object spectroscopy}},
  year          = {2024},
  month         = oct,
  pages         = {A288},
  volume        = {690},
  adsurl        = {https://ui.adsabs.harvard.edu/abs/2024A&A...690A.288B},
  archiveprefix = {arXiv},
  doi           = {10.1051/0004-6361/202347094},
  eid           = {A288},
  eprint        = {2306.02467},
  primaryclass  = {astro-ph.GA},
}

@Article{2023Natur.618..480R,
  author        = {{Roberts-Borsani}, Guido and {Treu}, Tommaso and {Chen}, Wenlei and {Morishita}, Takahiro and {Vanzella}, Eros and {Zitrin}, Adi and {Bergamini}, Pietro and {Castellano}, Marco and {Fontana}, Adriano and {Glazebrook}, Karl and {Grillo}, Claudio and {Kelly}, Patrick L. and {Merlin}, Emiliano and {Nanayakkara}, Themiya and {Paris}, Diego and {Rosati}, Piero and {Yang}, Lilan and {Acebron}, Ana and {Bonchi}, Andrea and {Boyett}, Kit and {Brada{\v{c}}}, Maru{\v{s}}a and {Brammer}, Gabriel and {Broadhurst}, Tom and {Calabr{\'o}}, Antonello and {Diego}, Jose M. and {Dressler}, Alan and {Furtak}, Lukas J. and {Filippenko}, Alexei V. and {Henry}, Alaina and {Koekemoer}, Anton M. and {Leethochawalit}, Nicha and {Malkan}, Matthew A. and {Mason}, Charlotte and {Mercurio}, Amata and {Metha}, Benjamin and {Pentericci}, Laura and {Pierel}, Justin and {Rieck}, Steven and {Roy}, Namrata and {Santini}, Paola and {Strait}, Victoria and {Strausbaugh}, Robert and {Trenti}, Michele and {Vulcani}, Benedetta and {Wang}, Lifan and {Wang}, Xin and {Windhorst}, Rogier A.},
  journal       = {\nat},
  title         = {{The nature of an ultra-faint galaxy in the cosmic dark ages seen with JWST}},
  year          = {2023},
  month         = jun,
  number        = {7965},
  pages         = {480-483},
  volume        = {618},
  adsurl        = {https://ui.adsabs.harvard.edu/abs/2023Natur.618..480R},
  archiveprefix = {arXiv},
  doi           = {10.1038/s41586-023-05994-w},
  eprint        = {2210.15639},
  primaryclass  = {astro-ph.GA},
}

@Article{2021AJ....162...47B,
  author        = {{Bouwens}, R.~J. and {Oesch}, P.~A. and {Stefanon}, M. and {Illingworth}, G. and {Labb{\'e}}, I. and {Reddy}, N. and {Atek}, H. and {Montes}, M. and {Naidu}, R. and {Nanayakkara}, T. and {Nelson}, E. and {Wilkins}, S.},
  journal       = {\aj},
  title         = {{New Determinations of the UV Luminosity Functions from z 9 to 2 Show a Remarkable Consistency with Halo Growth and a Constant Star Formation Efficiency}},
  year          = {2021},
  month         = aug,
  number        = {2},
  pages         = {47},
  volume        = {162},
  adsurl        = {https://ui.adsabs.harvard.edu/abs/2021AJ....162...47B},
  archiveprefix = {arXiv},
  doi           = {10.3847/1538-3881/abf83e},
  eid           = {47},
  eprint        = {2102.07775},
  primaryclass  = {astro-ph.GA},
}

@Article{2015ApJ...810...71F,
  author        = {{Finkelstein}, Steven L. and {Ryan}, Jr., Russell E. and {Papovich}, Casey and {Dickinson}, Mark and {Song}, Mimi and {Somerville}, Rachel S. and {Ferguson}, Henry C. and {Salmon}, Brett and {Giavalisco}, Mauro and {Koekemoer}, Anton M. and {Ashby}, Matthew L.~N. and {Behroozi}, Peter and {Castellano}, Marco and {Dunlop}, James S. and {Faber}, Sandy M. and {Fazio}, Giovanni G. and {Fontana}, Adriano and {Grogin}, Norman A. and {Hathi}, Nimish and {Jaacks}, Jason and {Kocevski}, Dale D. and {Livermore}, Rachael and {McLure}, Ross J. and {Merlin}, Emiliano and {Mobasher}, Bahram and {Newman}, Jeffrey A. and {Rafelski}, Marc and {Tilvi}, Vithal and {Willner}, S.~P.},
  journal       = {\apj},
  title         = {{The Evolution of the Galaxy Rest-frame Ultraviolet Luminosity Function over the First Two Billion Years}},
  year          = {2015},
  month         = sep,
  number        = {1},
  pages         = {71},
  volume        = {810},
  adsurl        = {https://ui.adsabs.harvard.edu/abs/2015ApJ...810...71F},
  archiveprefix = {arXiv},
  doi           = {10.1088/0004-637X/810/1/71},
  eid           = {71},
  eprint        = {1410.5439},
  primaryclass  = {astro-ph.GA},
}

@Article{2024MNRAS.533L..49Z,
  author        = {{Zhu}, Yongda and {Becker}, George D. and {Bosman}, Sarah E.~I. and {Cain}, Christopher and {Keating}, Laura C. and {Nasir}, Fahad and {D'Odorico}, Valentina and {Ba{\~n}ados}, Eduardo and {Bian}, Fuyan and {Bischetti}, Manuela and {Bolton}, James S. and {Chen}, Huanqing and {D'Aloisio}, Anson and {Davies}, Frederick B. and {Davies}, Rebecca L. and {Eilers}, Anna-Christina and {Fan}, Xiaohui and {Gaikwad}, Prakash and {Greig}, Bradley and {Haehnelt}, Martin G. and {Kulkarni}, Girish and {Lai}, Samuel and {Puchwein}, Ewald and {Qin}, Yuxiang and {Ryan-Weber}, Emma V. and {Satyavolu}, Sindhu and {Spina}, Benedetta and {Walter}, Fabian and {Wang}, Feige and {Wolfson}, Molly and {Yang}, Jinyi},
  journal       = {\mnras},
  title         = {{Damping wing-like features in the stacked Ly {\ensuremath{\alpha}} forest: Potential neutral hydrogen islands at z < 6}},
  year          = {2024},
  month         = sep,
  number        = {1},
  pages         = {L49-L56},
  volume        = {533},
  adsurl        = {https://ui.adsabs.harvard.edu/abs/2024MNRAS.533L..49Z},
  archiveprefix = {arXiv},
  doi           = {10.1093/mnrasl/slae061},
  eprint        = {2405.12275},
  primaryclass  = {astro-ph.CO},
}

@Article{2023ApJS..265....5H,
  author        = {{Harikane}, Yuichi and {Ouchi}, Masami and {Oguri}, Masamune and {Ono}, Yoshiaki and {Nakajima}, Kimihiko and {Isobe}, Yuki and {Umeda}, Hiroya and {Mawatari}, Ken and {Zhang}, Yechi},
  journal       = {\apjs},
  title         = {{A Comprehensive Study of Galaxies at z 9-16 Found in the Early JWST Data: Ultraviolet Luminosity Functions and Cosmic Star Formation History at the Pre-reionization Epoch}},
  year          = {2023},
  month         = mar,
  number        = {1},
  pages         = {5},
  volume        = {265},
  adsurl        = {https://ui.adsabs.harvard.edu/abs/2023ApJS..265....5H},
  archiveprefix = {arXiv},
  doi           = {10.3847/1538-4365/acaaa9},
  eid           = {5},
  eprint        = {2208.01612},
  primaryclass  = {astro-ph.GA},
}

@Article{2022ApJS..259...20H,
  author        = {{Harikane}, Yuichi and {Ono}, Yoshiaki and {Ouchi}, Masami and {Liu}, Chengze and {Sawicki}, Marcin and {Shibuya}, Takatoshi and {Behroozi}, Peter S. and {He}, Wanqiu and {Shimasaku}, Kazuhiro and {Arnouts}, Stephane and {Coupon}, Jean and {Fujimoto}, Seiji and {Gwyn}, Stephen and {Huang}, Jiasheng and {Inoue}, Akio K. and {Kashikawa}, Nobunari and {Komiyama}, Yutaka and {Matsuoka}, Yoshiki and {Willott}, Chris J.},
  journal       = {\apjs},
  title         = {{GOLDRUSH. IV. Luminosity Functions and Clustering Revealed with 4,000,000 Galaxies at z 2-7: Galaxy-AGN Transition, Star Formation Efficiency, and Implication for Evolution at z > 10}},
  year          = {2022},
  month         = mar,
  number        = {1},
  pages         = {20},
  volume        = {259},
  adsurl        = {https://ui.adsabs.harvard.edu/abs/2022ApJS..259...20H},
  archiveprefix = {arXiv},
  doi           = {10.3847/1538-4365/ac3dfc},
  eid           = {20},
  eprint        = {2108.01090},
  primaryclass  = {astro-ph.GA},
}

@Article{2016MNRAS.461.1100R,
  author        = {{Rowan-Robinson}, Michael and {Oliver}, Seb and {Wang}, Lingyu and {Farrah}, Duncan and {Clements}, David L. and {Gruppioni}, Carlotta and {Marchetti}, Lucia and {Rigopoulou}, Dimitra and {Vaccari}, Mattia},
  journal       = {\mnras},
  title         = {{The star formation rate density from z = 1 to 6}},
  year          = {2016},
  month         = sep,
  number        = {1},
  pages         = {1100-1111},
  volume        = {461},
  adsurl        = {https://ui.adsabs.harvard.edu/abs/2016MNRAS.461.1100R},
  archiveprefix = {arXiv},
  doi           = {10.1093/mnras/stw1169},
  eprint        = {1605.03937},
  primaryclass  = {astro-ph.GA},
}

@Article{2024Natur.626..975A,
  author        = {{Atek}, Hakim and {Labb{\'e}}, Ivo and {Furtak}, Lukas J. and {Chemerynska}, Iryna and {Fujimoto}, Seiji and {Setton}, David J. and {Miller}, Tim B. and {Oesch}, Pascal and {Bezanson}, Rachel and {Price}, Sedona H. and {Dayal}, Pratika and {Zitrin}, Adi and {Kokorev}, Vasily and {Weaver}, John R. and {Brammer}, Gabriel and {Dokkum}, Pieter van and {Williams}, Christina C. and {Cutler}, Sam E. and {Feldmann}, Robert and {Fudamoto}, Yoshinobu and {Greene}, Jenny E. and {Leja}, Joel and {Maseda}, Michael V. and {Muzzin}, Adam and {Pan}, Richard and {Papovich}, Casey and {Nelson}, Erica J. and {Nanayakkara}, Themiya and {Stark}, Daniel P. and {Stefanon}, Mauro and {Suess}, Katherine A. and {Wang}, Bingjie and {Whitaker}, Katherine E.},
  journal       = {\nat},
  title         = {{Most of the photons that reionized the Universe came from dwarf galaxies}},
  year          = {2024},
  month         = feb,
  number        = {8001},
  pages         = {975-978},
  volume        = {626},
  adsurl        = {https://ui.adsabs.harvard.edu/abs/2024Natur.626..975A},
  archiveprefix = {arXiv},
  doi           = {10.1038/s41586-024-07043-6},
  eprint        = {2308.08540},
  primaryclass  = {astro-ph.GA},
}

@Article{2008A&A...488..463M,
  author        = {{Maiolino}, R. and {Nagao}, T. and {Grazian}, A. and {Cocchia}, F. and {Marconi}, A. and {Mannucci}, F. and {Cimatti}, A. and {Pipino}, A. and {Ballero}, S. and {Calura}, F. and {Chiappini}, C. and {Fontana}, A. and {Granato}, G.~L. and {Matteucci}, F. and {Pastorini}, G. and {Pentericci}, L. and {Risaliti}, G. and {Salvati}, M. and {Silva}, L.},
  journal       = {\aap},
  title         = {{AMAZE. I. The evolution of the mass-metallicity relation at z > 3}},
  year          = {2008},
  month         = sep,
  number        = {2},
  pages         = {463-479},
  volume        = {488},
  adsurl        = {https://ui.adsabs.harvard.edu/abs/2008A&A...488..463M},
  archiveprefix = {arXiv},
  doi           = {10.1051/0004-6361:200809678},
  eprint        = {0806.2410},
  primaryclass  = {astro-ph},
}

@Article{2024ApJ...966..133S,
  author        = {{Sears}, Huei and {Chornock}, Ryan and {Strader}, Jay and {Perley}, Daniel A. and {Blanchard}, Peter K. and {Margutti}, Raffaella and {Tanvir}, Nial R.},
  journal       = {\apj},
  title         = {{Constraints on the z {\ensuremath{\sim}} 5 Star-forming Galaxy Luminosity Function From Hubble Space Telescope Imaging of an Unbiased and Complete Sample of Long Gamma-Ray Burst Host Galaxies}},
  year          = {2024},
  month         = may,
  number        = {1},
  pages         = {133},
  volume        = {966},
  adsurl        = {https://ui.adsabs.harvard.edu/abs/2024ApJ...966..133S},
  archiveprefix = {arXiv},
  doi           = {10.3847/1538-4357/ad2e93},
  eid           = {133},
  eprint        = {2308.14248},
  primaryclass  = {astro-ph.GA},
}

@Article{2013arXiv1305.1630K,
  author        = {{Kistler}, Matthew D. and {Yuksel}, Hasan and {Hopkins}, Andrew M.},
  journal       = {arXiv e-prints},
  title         = {{The Cosmic Star Formation Rate from the Faintest Galaxies in the Unobservable Universe}},
  year          = {2013},
  month         = may,
  pages         = {arXiv:1305.1630},
  adsurl        = {https://ui.adsabs.harvard.edu/abs/2013arXiv1305.1630K},
  archiveprefix = {arXiv},
  doi           = {10.48550/arXiv.1305.1630},
  eid           = {arXiv:1305.1630},
  eprint        = {1305.1630},
  primaryclass  = {astro-ph.CO},
}

@Article{1999ApJ...521...64M,
  author        = {{Meurer}, Gerhardt R. and {Heckman}, Timothy M. and {Calzetti}, Daniela},
  journal       = {\apj},
  title         = {{Dust Absorption and the Ultraviolet Luminosity Density at z \raisebox{-0.5ex}\textasciitilde 3 as Calibrated by Local Starburst Galaxies}},
  year          = {1999},
  month         = aug,
  number        = {1},
  pages         = {64-80},
  volume        = {521},
  adsurl        = {https://ui.adsabs.harvard.edu/abs/1999ApJ...521...64M},
  archiveprefix = {arXiv},
  doi           = {10.1086/307523},
  eprint        = {astro-ph/9903054},
  primaryclass  = {astro-ph},
}

@Article{2014ApJ...793..115B,
  author        = {{Bouwens}, R.~J. and {Illingworth}, G.~D. and {Oesch}, P.~A. and {Labb{\'e}}, I. and {van Dokkum}, P.~G. and {Trenti}, M. and {Franx}, M. and {Smit}, R. and {Gonzalez}, V. and {Magee}, D.},
  journal       = {\apj},
  title         = {{UV-continuum Slopes of >4000 z \raisebox{-0.5ex}\textasciitilde 4-8 Galaxies from the HUDF/XDF, HUDF09, ERS, CANDELS-South, and CANDELS-North Fields}},
  year          = {2014},
  month         = oct,
  number        = {2},
  pages         = {115},
  volume        = {793},
  adsurl        = {https://ui.adsabs.harvard.edu/abs/2014ApJ...793..115B},
  archiveprefix = {arXiv},
  doi           = {10.1088/0004-637X/793/2/115},
  eid           = {115},
  eprint        = {1306.2950},
  primaryclass  = {astro-ph.CO},
}

@Article{2013ApJ...768L..37T,
  author        = {{Tacchella}, Sandro and {Trenti}, Michele and {Carollo}, C. Marcella},
  journal       = {\apjl},
  title         = {{A Physical Model for the 0 <\raisebox{-0.5ex}\textasciitilde z <\raisebox{-0.5ex}\textasciitilde 8 Redshift Evolution of the Galaxy Ultraviolet Luminosity and Stellar Mass Functions}},
  year          = {2013},
  month         = may,
  number        = {2},
  pages         = {L37},
  volume        = {768},
  adsurl        = {https://ui.adsabs.harvard.edu/abs/2013ApJ...768L..37T},
  archiveprefix = {arXiv},
  doi           = {10.1088/2041-8205/768/2/L37},
  eid           = {L37},
  eprint        = {1211.2825},
  primaryclass  = {astro-ph.CO},
}

@Article{2012MNRAS.423..862K,
  author        = {{Kuhlen}, Michael and {Faucher-Gigu{\`e}re}, Claude-Andr{\'e}},
  journal       = {\mnras},
  title         = {{Concordance models of reionization: implications for faint galaxies and escape fraction evolution}},
  year          = {2012},
  month         = jun,
  number        = {1},
  pages         = {862-876},
  volume        = {423},
  adsurl        = {https://ui.adsabs.harvard.edu/abs/2012MNRAS.423..862K},
  archiveprefix = {arXiv},
  doi           = {10.1111/j.1365-2966.2012.20924.x},
  eprint        = {1201.0757},
  primaryclass  = {astro-ph.CO},
}

@Article{2019MNRAS.486.3805B,
  author        = {{Bhatawdekar}, Rachana and {Conselice}, Christopher J. and {Margalef-Bentabol}, Berta and {Duncan}, Kenneth},
  journal       = {\mnras},
  title         = {{Evolution of the galaxy stellar mass functions and UV luminosity functions at z = 6-9 in the Hubble Frontier Fields}},
  year          = {2019},
  month         = jul,
  number        = {3},
  pages         = {3805-3830},
  volume        = {486},
  adsurl        = {https://ui.adsabs.harvard.edu/abs/2019MNRAS.486.3805B},
  archiveprefix = {arXiv},
  doi           = {10.1093/mnras/stz866},
  eprint        = {1807.07580},
  primaryclass  = {astro-ph.GA},
}

@Article{2019MNRAS.488..419W,
  author        = {{Wu}, Xiaohan and {Kannan}, Rahul and {Marinacci}, Federico and {Vogelsberger}, Mark and {Hernquist}, Lars},
  journal       = {\mnras},
  title         = {{Simulating the effect of photoheating feedback during reionization}},
  year          = {2019},
  month         = sep,
  number        = {1},
  pages         = {419-437},
  volume        = {488},
  adsurl        = {https://ui.adsabs.harvard.edu/abs/2019MNRAS.488..419W},
  archiveprefix = {arXiv},
  doi           = {10.1093/mnras/stz1726},
  eprint        = {1903.06167},
  primaryclass  = {astro-ph.GA},
}

@Article{2021MNRAS.507.6108O,
  author        = {{Ocvirk}, Pierre and {Lewis}, Joseph S.~W. and {Gillet}, Nicolas and {Chardin}, Jonathan and {Aubert}, Dominique and {Deparis}, Nicolas and {Th{\'e}lie}, {\'E}milie},
  journal       = {\mnras},
  title         = {{Lyman-alpha opacities at z = 4-6 require low mass, radiatively-suppressed galaxies to drive cosmic reionization}},
  year          = {2021},
  month         = nov,
  number        = {4},
  pages         = {6108-6117},
  volume        = {507},
  adsurl        = {https://ui.adsabs.harvard.edu/abs/2021MNRAS.507.6108O},
  archiveprefix = {arXiv},
  doi           = {10.1093/mnras/stab2502},
  eprint        = {2105.01663},
  primaryclass  = {astro-ph.CO},
}

@Article{2025SCPMA..6839501Y,
  author        = {{Yuan}, Weimin and {Dai}, Lixin and {Feng}, Hua and {Jin}, Chichuan and {Jonker}, Peter and {Kuulkers}, Erik and {Liu}, Yuan and {Nandra}, Kirpal and {O'Brien}, Paul and {Piro}, Luigi and {Rau}, Arne and {Rea}, Nanda and {Sanders}, Jeremy and {Tao}, Lian and {Wang}, Junfeng and {Wu}, Xuefeng and {Zhang}, Bing and {Zhang}, Shuangnan and {Ai}, Shunke and {Buchner}, Johannes and {Bulbul}, Esra and {Chen}, Hechao and {Chen}, Minghua and {Chen}, Yong and {Chen}, Yu-Peng and {Coleiro}, Alexis and {Zelati}, Francesco Coti and {Dai}, Zigao and {Fan}, Xilong and {Fan}, Zhou and {Friedrich}, Susanne and {Gao}, He and {Ge}, Chong and {Ge}, Mingyu and {Geng}, Jinjun and {Ghirlanda}, Giancarlo and {Gianfagna}, Giulia and {Gou}, Lijun and {Guillot}, S{\'e}bastien and {Hou}, Xian and {Hu}, Jingwei and {Huang}, Yongfeng and {Ji}, Long and {Jia}, Shumei and {Komossa}, S. and {Kong}, Albert K.~H. and {Lan}, Lin and {Li}, An and {Li}, Ang and {Li}, Chengkui and {Li}, Dongyue and {Li}, Jian and {Li}, Zhaosheng and {Ling}, Zhixing and {Liu}, Ang and {Liu}, Jinzhong and {Liu}, Liangduan and {Liu}, Zhu and {Luo}, Jiawei and {Ma}, Ruican and {Maggi}, Pierre and {Maitra}, Chandreyee and {Marino}, Alessio and {Ng}, Stephen Chi-Yung and {Pan}, Haiwu and {Rukdee}, Surangkhana and {Soria}, Roberto and {Sun}, Hui and {Tam}, Pak-Hin Thomas and {Thakur}, Aishwarya Linesh and {Tian}, Hui and {Troja}, Eleonora and {Wang}, Wei and {Wang}, Xiangyu and {Wang}, Yanan and {Wei}, Junjie and {Wen}, Sixiang and {Wu}, Jianfeng and {Wu}, Ting and {Xiao}, Di and {Xu}, Dong and {Xu}, Renxin and {Xu}, Yanjun and {Xu}, Yu and {Yang}, Haonan and {You}, Bei and {Yu}, Heng and {Yu}, Yunwei and {Zhang}, Binbin and {Zhang}, Chen and {Zhang}, Guobao and {Zhang}, Liang and {Zhang}, Wenda and {Zhang}, Yu and {Zhou}, Ping and {Zou}, Zecheng},
  journal       = {Science China Physics, Mechanics, and Astronomy},
  title         = {{Science objectives of the Einstein Probe mission}},
  year          = {2025},
  month         = mar,
  number        = {3},
  pages         = {239501},
  volume        = {68},
  adsurl        = {https://ui.adsabs.harvard.edu/abs/2025SCPMA..6839501Y},
  archiveprefix = {arXiv},
  doi           = {10.1007/s11433-024-2600-3},
  eid           = {239501},
  eprint        = {2501.07362},
  primaryclass  = {astro-ph.HE},
}

@Article{2016arXiv161006892W,
  author        = {{Wei}, J. and {Cordier}, B. and {Antier}, S. and {Antilogus}, P. and {Atteia}, J. -L. and {Bajat}, A. and {Basa}, S. and {Beckmann}, V. and {Bernardini}, M.~G. and {Boissier}, S. and {Bouchet}, L. and {Burwitz}, V. and {Claret}, A. and {Dai}, Z. -G. and {Daigne}, F. and {Deng}, J. and {Dornic}, D. and {Feng}, H. and {Foglizzo}, T. and {Gao}, H. and {Gehrels}, N. and {Godet}, O. and {Goldwurm}, A. and {Gonzalez}, F. and {Gosset}, L. and {G{\"o}tz}, D. and {Gouiffes}, C. and {Grise}, F. and {Gros}, A. and {Guilet}, J. and {Han}, X. and {Huang}, M. and {Huang}, Y. -F. and {Jouret}, M. and {Klotz}, A. and {La Marle}, O. and {Lachaud}, C. and {Le Floch}, E. and {Lee}, W. and {Leroy}, N. and {Li}, L. -X. and {Li}, S.~C. and {Li}, Z. and {Liang}, E. -W. and {Lyu}, H. and {Mercier}, K. and {Migliori}, G. and {Mochkovitch}, R. and {O'Brien}, P. and {Osborne}, J. and {Paul}, J. and {Perinati}, E. and {Petitjean}, P. and {Piron}, F. and {Qiu}, Y. and {Rau}, A. and {Rodriguez}, J. and {Schanne}, S. and {Tanvir}, N. and {Vangioni}, E. and {Vergani}, S. and {Wang}, F. -Y. and {Wang}, J. and {Wang}, X. -G. and {Wang}, X. -Y. and {Watson}, A. and {Webb}, N. and {Wei}, J.~J. and {Willingale}, R. and {Wu}, C. and {Wu}, X. -F. and {Xin}, L. -P. and {Xu}, D. and {Yu}, S. and {Yu}, W. -F. and {Yu}, Y. -W. and {Zhang}, B. and {Zhang}, S. -N. and {Zhang}, Y. and {Zhou}, X.~L.},
  journal       = {arXiv e-prints},
  title         = {{The Deep and Transient Universe in the SVOM Era: New Challenges and Opportunities - Scientific prospects of the SVOM mission}},
  year          = {2016},
  month         = oct,
  pages         = {arXiv:1610.06892},
  adsurl        = {https://ui.adsabs.harvard.edu/abs/2016arXiv161006892W},
  archiveprefix = {arXiv},
  doi           = {10.48550/arXiv.1610.06892},
  eid           = {arXiv:1610.06892},
  eprint        = {1610.06892},
  primaryclass  = {astro-ph.IM},
}

@Article{2021ExA....52..219T,
  author        = {{Tanvir}, N.~R. and {Le Floc'h}, E. and {Christensen}, L. and {Caruana}, J. and {Salvaterra}, R. and {Ghirlanda}, G. and {Ciardi}, B. and {Maio}, U. and {D'Odorico}, V. and {Piedipalumbo}, E. and {Campana}, S. and {Noterdaeme}, P. and {Graziani}, L. and {Amati}, L. and {Bagoly}, Z. and {Bal{\'a}zs}, L.~G. and {Basa}, S. and {Behar}, E. and {De Cia}, A. and {Della Valle}, M. and {De Pasquale}, M. and {Frontera}, F. and {Gomboc}, A. and {G{\"o}tz}, D. and {Horvath}, I. and {Hudec}, R. and {Mereghetti}, S. and {O'Brien}, P.~T. and {Osborne}, J.~P. and {Paltani}, S. and {Rosati}, P. and {Sergijenko}, O. and {Stanway}, E.~R. and {Sz{\'e}csi}, D. and {Toth}, L.~V. and {Urata}, Y. and {Vergani}, S. and {Zane}, S.},
  journal       = {Experimental Astronomy},
  title         = {{Exploration of the high-redshift universe enabled by THESEUS}},
  year          = {2021},
  month         = dec,
  number        = {3},
  pages         = {219-244},
  volume        = {52},
  adsurl        = {https://ui.adsabs.harvard.edu/abs/2021ExA....52..219T},
  archiveprefix = {arXiv},
  doi           = {10.1007/s10686-021-09778-w},
  eprint        = {2104.09532},
  primaryclass  = {astro-ph.IM},
}

@Article{2015MNRAS.447..499M,
  author        = {{McGreer}, Ian D. and {Mesinger}, Andrei and {D'Odorico}, Valentina},
  journal       = {\mnras},
  title         = {{Model-independent evidence in favour of an end to reionization by z {\ensuremath{\approx}} 6}},
  year          = {2015},
  month         = feb,
  number        = {1},
  pages         = {499-505},
  volume        = {447},
  adsurl        = {https://ui.adsabs.harvard.edu/abs/2015MNRAS.447..499M},
  archiveprefix = {arXiv},
  doi           = {10.1093/mnras/stu2449},
  eprint        = {1411.5375},
  primaryclass  = {astro-ph.CO},
}

@Article{2018PASJ...70S..13O,
  author        = {{Ouchi}, Masami and {Harikane}, Yuichi and {Shibuya}, Takatoshi and {Shimasaku}, Kazuhiro and {Taniguchi}, Yoshiaki and {Konno}, Akira and {Kobayashi}, Masakazu and {Kajisawa}, Masaru and {Nagao}, Tohru and {Ono}, Yoshiaki and {Inoue}, Akio K. and {Umemura}, Masayuki and {Mori}, Masao and {Hasegawa}, Kenji and {Higuchi}, Ryo and {Komiyama}, Yutaka and {Matsuda}, Yuichi and {Nakajima}, Kimihiko and {Saito}, Tomoki and {Wang}, Shiang-Yu},
  journal       = {\pasj},
  title         = {{Systematic Identification of LAEs for Visible Exploration and Reionization Research Using Subaru HSC (SILVERRUSH). I. Program strategy and clustering properties of {\ensuremath{\sim}}2000 Ly{\ensuremath{\alpha}} emitters at z = 6-7 over the 0.3-0.5 Gpc$^{2}$ survey area}},
  year          = {2018},
  month         = jan,
  pages         = {S13},
  volume        = {70},
  adsurl        = {https://ui.adsabs.harvard.edu/abs/2018PASJ...70S..13O},
  archiveprefix = {arXiv},
  doi           = {10.1093/pasj/psx074},
  eid           = {S13},
  eprint        = {1704.07455},
  primaryclass  = {astro-ph.GA},
}

@Article{2021ApJ...919..120M,
  author        = {{Morales}, Alexa M. and {Mason}, Charlotte A. and {Bruton}, Sean and {Gronke}, Max and {Haardt}, Francesco and {Scarlata}, Claudia},
  journal       = {\apj},
  title         = {{The Evolution of the Lyman-alpha Luminosity Function during Reionization}},
  year          = {2021},
  month         = oct,
  number        = {2},
  pages         = {120},
  volume        = {919},
  adsurl        = {https://ui.adsabs.harvard.edu/abs/2021ApJ...919..120M},
  archiveprefix = {arXiv},
  doi           = {10.3847/1538-4357/ac1104},
  eid           = {120},
  eprint        = {2101.01205},
  primaryclass  = {astro-ph.GA},
}

\clearpage

\begin{figure}
\includegraphics[width=.6\linewidth]{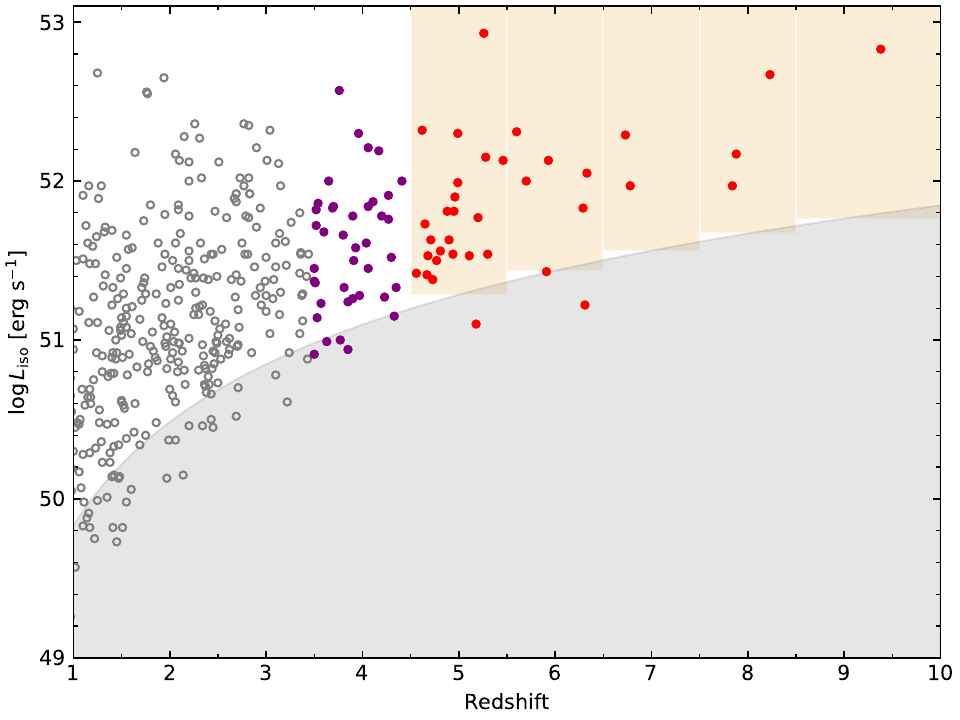}
\caption{The luminosity-redshift distribution of Swift LGRBs.
The gray shaded region represents an approximate luminosity threshold for detection.
The purple circles are the 39 LGRBs at $3.5\leqslant z<4.5$ used as our benchmark.
The 36 bursts at $z\geqslant4.5$ are separated into different redshift bins (orange shaded bands) used to estimate
the high-redshift SFRD.
\label{fig:lisoz}}
\end{figure}

\begin{figure}
\includegraphics[width=.6\linewidth]{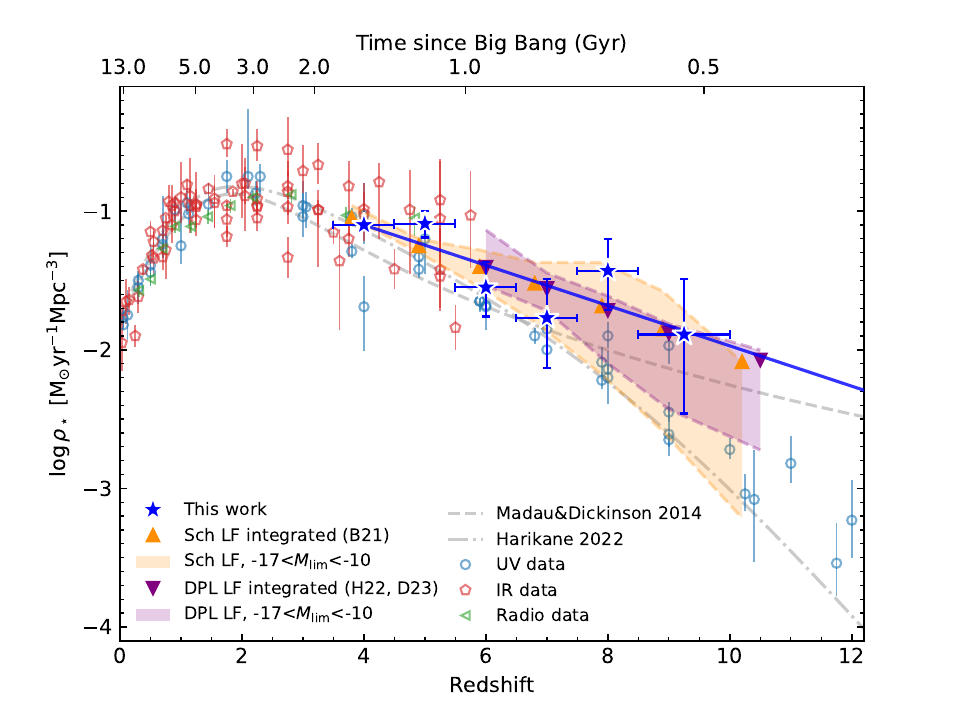}
\caption{Cosmic star formation history.
The blue stars represent our estimates from Swift LGRBs, with the
blue solid line showing the log-linear fit to these points ($k=-0.15$).
The orange shaded region shows the result of integrating the Schechter LFs
of \citet{2021AJ....162...47B} adopting the limits from $M_\mathrm{lim}=-17$ to $-10$,
while the purple shaded region shows that from the DPL fits \citep{2022ApJS..259...20H,2023MNRAS.518.6011D}.
In particular, the orange up triangles denote the results for the Schechter model
down to $M_\mathrm{lim}=-17,\,-15,\,-14,\,-13,\,-13,\,-12,\,-10$ for $\left\langle z\right\rangle =3.8$,
4.9, 5.9, 6.8, 7.9, 8.9, 10.2, respectively. The purple down triangles
denote the results for the DPL models with $M_\mathrm{lim}=-15,\,-14,\,-12,\,-11,\,-11$ for $\left\langle z\right\rangle =6$,
7, 8, 9, 10.5, respectively.
The grey dashed and dot-dashed lines denote the empirical fits from
\citet{2014ARA&A..52..415M} and \citet{2022ApJS..259...20H}, respectively.
The open circles, pentagons and left triangles represent observations determined from the UV data
\citep{2014ARA&A..52..415M,2015ApJ...810...71F,2019MNRAS.486.3805B,2020ApJ...902..112B,2023ApJS..265....5H,2024MNRAS.533.3222D,2024ApJ...966...74W},
IR/(sub)millimeter data \citep{2014ARA&A..52..415M,2016MNRAS.461.1100R,2018ApJ...853..172L,2020A&A...643A...8G,2024A&A...681A.118T,2026A&A...705A.255T},
and radio data \citep{2017A&A...602A...5N}, respectively. All UV values shown here are determined using a limit of $M_\mathrm{lim}=17$,
with the only exception of \citet{2019MNRAS.486.3805B} using $M_\mathrm{lim}=13.5$.
All results have been converted to the Salpeter IMF if necessary.
\label{fig:sfrd}}
\end{figure}

\begin{figure}
\includegraphics[width=1.0\linewidth]{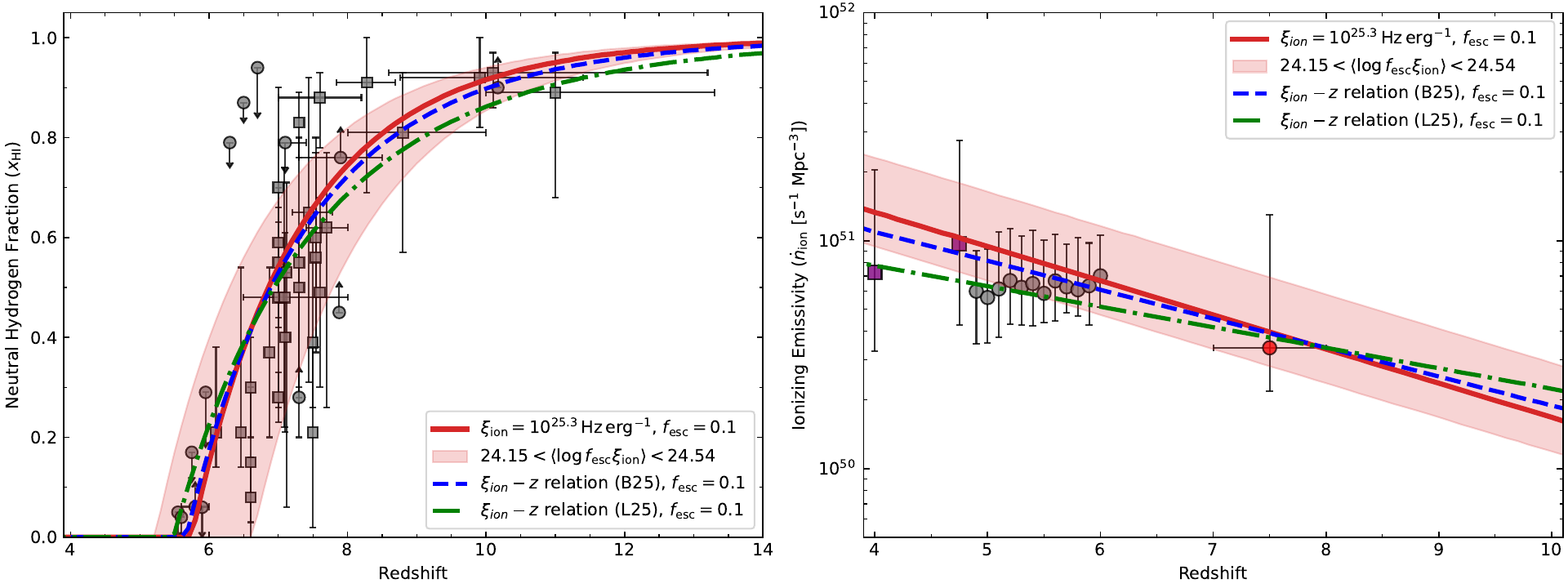}
\caption{Hydrogen reionization history.
Left panel: The predictions of the timeline of the neutral hydrogen fraction $x_\mathrm{HI}$, compared
with observational constraints derived from galaxies and quasars \citep{2014ApJ...797...16K,2015MNRAS.447..499M,
2017MNRAS.466.4239G,2019MNRAS.484.5094G,2018Natur.553..473B,2018ApJ...864..142D,2018PASJ...70...55I,2018ApJ...856....2M,
2019MNRAS.485.3947M,2018PASJ...70S..13O,2019ApJ...878...12H,2020ApJ...904..144J,2020ApJ...896...23W,2020MNRAS.495.3602W,
2020ApJ...897L..14Y,2021ApJ...923..229G,2021ApJ...919..120M,2022ApJ...926..230N,2022ApJ...932...76Z,2024MNRAS.533L..49Z,
2023ApJ...942...59J,2023ApJ...947L..24M,2024ApJ...969..162D,2024ApJ...973....8H,2024ApJ...967...28N,2024ApJ...975..208T,2024ApJ...971..124U}.
Right panel: The ionizing emissivity $\dot{n}_\mathrm{ion}$ as a function of redshift,
compared with observations \citep{2013MNRAS.436.1023B,2023MNRAS.525.4093G,2024ApJ...969...12R}.
The red solid line represents the timeline derived from our fiducial values.
The blue dashed line represents the result for a redshift-dependent
$\xi_\mathrm{ion}$ based on the JWST measurements by \citet{2025MNRAS.537.3245B}:
$\log (\xi_\mathrm{ion})=(0.023\pm 0.005)z+(25.12\pm 0.03)$, while
the green dot-dashed line shows the result from another similar measurements of \citet{2025A&A...698A.302L}:
$\log (\xi_\mathrm{ion})=(0.006\pm 0.012)z+(24.82\pm 0.07)$.
The red shaded region represents the values of
$24.15<\langle\log f_\mathrm{esc}\xi_\mathrm{ion}\rangle<24.54$,
constrained by the errors of the measured CMB optical depth
by the Planck satellite (see Figure \ref{fig:tau}).
\label{fig:EoR}}
\end{figure}

\begin{figure}
\includegraphics[width=.6\linewidth]{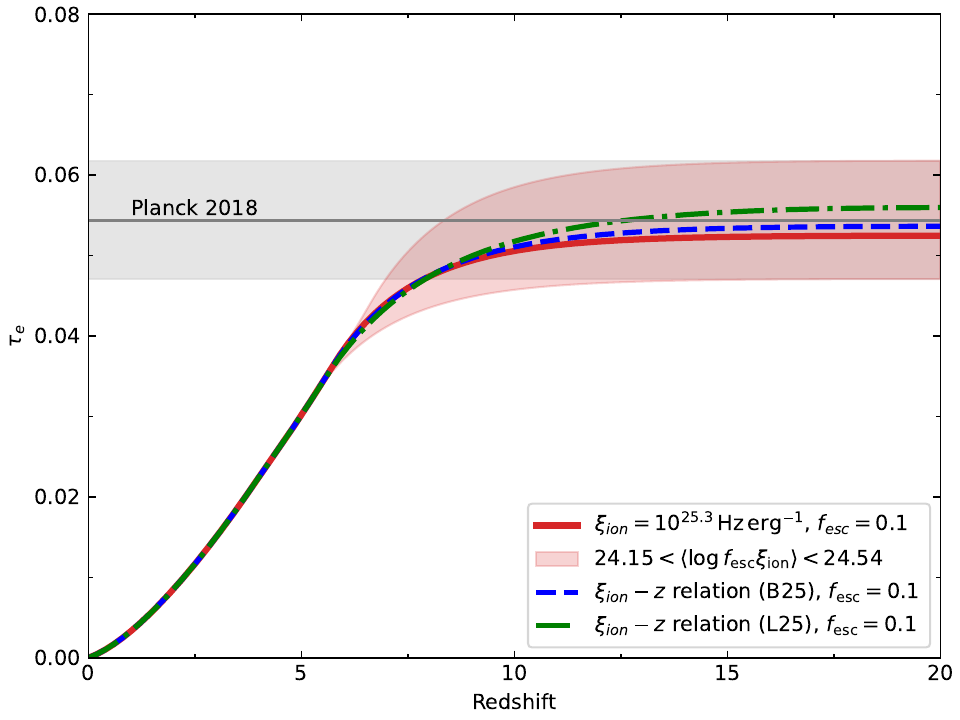}
\caption{Predictions of the Thomson scattering optical depth.
The gray line and region represent the measurement provided by Planck,
i.e. $\tau_{e}=0.0544\pm0.0073$.
The red solid, blue dashed, and green dot-dashed lines correspond to the results
shown in Figure \ref{fig:EoR}. The red shaded region represents
the values of $24.15<\langle\log f_\mathrm{esc}\xi_\mathrm{ion}\rangle<24.54$
constrained by the measured errors.
\label{fig:tau}}
\end{figure}

\begin{figure}
\includegraphics[width=.6\linewidth]{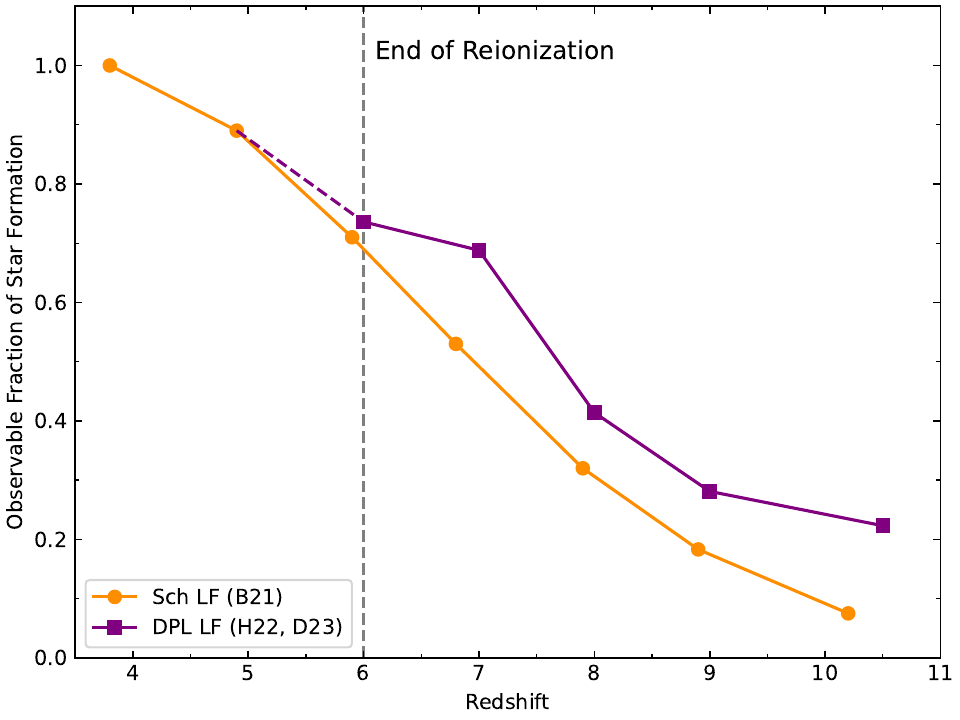}
\caption{Evolution of the observable fraction of star formation, $f_{\mathrm{obs}}(z)=\dot{\rho}_{\star}^{\mathrm{obs}}(z)/\dot{\rho}_{\star}^{\mathrm{tot}}(z)$,
where $\dot{\rho}_{\star}^{\mathrm{obs}}(z)$ and $\dot{\rho}_{\star}^{\mathrm{tot}}(z)$ represent the SFRDs
in observable galaxies with $M_{\mathrm{UV}}\leqslant-17$ and all galaxies, respectively. Shown are the results integrated from the two
LF models (Schechter as the orange line, and DPL as the purple line).
\label{fig:fraction}}
\end{figure}

\clearpage

\startlongtable
\begin{deluxetable*}{lcccccc}
\tabletypesize{\scriptsize}
\tablewidth{0pt}
\tablecaption{List of LGRBs with $z\geqslant3.5$\label{tab1}}
\tablehead{
\colhead{GRB} & \colhead{$z$} & \colhead{Ref} & \colhead{$T_{90}$} & \colhead{$S_{\gamma}$} &
\colhead{$\Gamma$} & \colhead{$\log(L_{\mathrm{iso}})$}\\
 & & & \colhead{(s)} & \colhead{($10^{-7}\,\mathrm{erg\, cm^{-2}}$)} & &
 \colhead{($\mathrm{erg\, s^{-1}}$)}}
\startdata
GRB251126A & 3.52 & 1 & 9 & 4.4 & 1.76 & 51.72\\
GRB251017A & 4.327 & 2 & 116.9 & 12 & 1.4 & 51.15\\
GRB250925A & 3.899 & 3 & 64.51 & 27 & 1.87 & 51.78\\
GRB250725A & 5.26 & 4 & 9.58 & 50 & 1.15 & 52.93\\
GRB250114A & 4.732 & 5 & 294 & 34.1 & 1.77 & 51.38\\
GRB241025A & 4.2 & 6 & 130.09 & 60.58 & 1.37 & 51.78\\
GRB240419A & 5.178 & 7 & 114.26 & 6.28 & 1.64 & 51.1\\
GRB240218A & 6.782 & 8 & 258.03 & 61.88 & 1.62 & 51.97\\
GRB230506C & 3.7 & 9 & 31 & 16.88 & 1.88 & 51.84\\
GRB230414B & 3.568 & 10 & 29.06 & 4.72 & 1.6 & 51.23\\
GRB230116D & 3.81 & 11 & 41 & 8.13 & 1.38 & 51.33\\
GRB221110A & 4.06 & 12 & 8.98 & 5.12 & 1.41 & 51.84\\
GRB220521A & 5.6 & 13 & 13.55 & 8.05 & 1.97 & 52.31\\
GRB220117A & 4.961 & 14 & 50.56 & 16.75 & 1.86 & 51.91\\
GRB220101A & 4.618 & 15 & 161.89 & 236.93 & 1.26 & 52.32\\
GRB210905A & 6.3118 & 16 & 778.4 & 42.82 & 1.51 & 51.22\\
GRB210610A & 3.54 & 17 & 13.62 & 10.41 & 1.41 & 51.86\\
GRB201221A & 5.7 & 18 & 44.32 & 19.23 & 1.4 & 52\\
GRB201014A & 4.56 & 19 & 36.2 & 3.07 & 2.55 & 51.42\\
GRB191004B & 3.503 & 20 & 300.13 & 30.11 & 1.1 & 50.91\\
GRB170405A & 3.51 & 21 & 165.31 & 38.77 & 1.51 & 51.36\\
GRB170202A & 3.645 & 22 & 37.76 & 33.38 & 1.67 & 52\\
GRB160327A & 4.99 & 23 & 33.74 & 13.65 & 1.84 & 51.99\\
GRB160203A & 3.518 & 22 & 17.44 & 9.93 & 1.94 & 51.82\\
GRB151112A & 4.27 & 24 & 19.32 & 9.4 & 1.77 & 51.91\\
GRB151111A & 3.5 & 25 & 76.93 & 17 & 1.76 & 51.37\\
GRB151027B & 4.063 & 22 & 80 & 14.73 & 1.82 & 51.45\\
GRB150120B & 3.5 & 26 & 24.34 & 5.34 & 2.24 & 51.45\\
GRB140614A & 4.233 & 22 & 77.39 & 10.35 & 1.5 & 51.27\\
GRB140518A & 4.707 & 27 & 60.52 & 11.69 & 1.89 & 51.63\\
GRB140515A & 6.327 & 28 & 23.42 & 6.72 & 1.78 & 52.05\\
GRB140428A & 4.68 & 24 & 17.42 & 3.4 & 1.54 & 51.53\\
GRB140419A & 3.961 & 29 & 80.08 & 157.28 & 1.15 & 52.3\\
GRB140311A & 4.954 & 22 & 70.48 & 20.97 & 1.7 & 51.81\\
GRB140304A & 5.283 & 30 & 14.78 & 11.28 & 1.3 & 52.15\\
GRB131117A & 4.042 & 22 & 10.88 & 2.98 & 1.81 & 51.61\\
GRB130606A & 5.9134 & 31 & 276.66 & 27.57 & 1.52 & 51.43\\
GRB130514A & 3.6 & 32 & 214.19 & 89.96 & 1.76 & 51.68\\
GRB130408A & 3.758 & 22 & 4.24 & 16.71 & 1.15 & 52.57\\
GRB120923A & 7.84 & 33 & 26.08 & 3.89 & 1.8 & 51.97\\
GRB120909A & 3.929 & 22 & 220.6 & 75 & 1.37 & 51.58\\
GRB120805A & 3.9 & 22 & 48 & 8.41 & 1.21 & 51.26\\
GRB120802A & 3.796 & 34 & 50.29 & 17.38 & 1.85 & 51.66\\
GRB120712A & 4.175 & 22 & 14.81 & 18.44 & 1.36 & 52.2\\
GRB120521C & 5.93 & 35 & 27.07 & 12.02 & 1.66 & 52.13\\
GRB111008A & 4.99 & 22 & 62.85 & 51.99 & 1.82 & 52.3\\
GRB100905A & 7.88 & 24 & 3.4 & 1.73 & 1.09 & 52.17\\
GRB100513A & 4.77 & 21 & 83.5 & 14.29 & 1.59 & 51.5\\
GRB100302A & 4.81 & 21 & 17.95 & 3.16 & 1.72 & 51.56\\
GRB100219A & 4.66723 & 36 & 27.57 & 4.65 & 1.33 & 51.41\\
GRB090519 & 3.85 & 21 & 58.04 & 11.8 & 0.85 & 51.24\\
GRB090516A & 4.11 & 37 & 181.01 & 92.15 & 1.67 & 51.87\\
GRB090429B & 9.38 & 38 & 5.58 & 3.73 & 1.89 & 52.83\\
GRB090423 & 8.23 & 39 & 10.3 & 6.85 & 1.81 & 52.67\\
GRB090205 & 4.65 & 40 & 8.81 & 2.03 & 2.01 & 51.73\\
GRB081029 & 3.85 & 21 & 275.1 & 21.45 & 1.41 & 50.94\\
GRB080913 & 6.733 & 41 & 7.46 & 5.72 & 1.2 & 52.29\\
GRB080129 & 4.35 & 21 & 50.18 & 8.39 & 1.27 & 51.33\\
GRB071025 & 4.88 & 24 & 241.3 & 73.16 & 1.71 & 51.81\\
GRB070721B & 3.6298 & 42 & 336.86 & 35.64 & 1.24 & 50.99\\
GRB060927 & 5.4636 & 42 & 22.42 & 12.07 & 1.61 & 52.13\\
GRB060906 & 3.6856 & 42 & 44.59 & 22.86 & 1.98 & 51.83\\
GRB060605 & 3.773 & 43 & 79.84 & 7.11 & 1.52 & 51\\
GRB060522 & 5.11 & 21 & 69.12 & 11.29 & 1.54 & 51.53\\
GRB060510B & 4.941 & 44 & 262.94 & 39.81 & 1.77 & 51.54\\
GRB060223A & 4.41 & 45 & 11.32 & 6.74 & 1.69 & 52\\
GRB060210 & 3.9122 & 42 & 288 & 76.81 & 1.48 & 51.5\\
GRB060206 & 4.0559 & 42 & 7.55 & 8.81 & 1.66 & 52.21\\
GRB060115 & 3.5328 & 42 & 139.09 & 18.27 & 1.7 & 51.14\\
GRB050922B & 4.9 & 46 & 157.02 & 23.72 & 2.11 & 51.63\\
GRB050904 & 6.295 & 47 & 181.58 & 52.08 & 1.23 & 51.83\\
GRB050814 & 5.3 & 48 & 142.85 & 19.07 & 1.75 & 51.54\\
GRB050803 & 4.3 & 46 & 88.12 & 21.64 & 1.39 & 51.52\\
GRB050730 & 3.9693 & 42 & 154.6 & 23.63 & 1.51 & 51.28\\
GRB050505 & 4.2748 & 49 & 58.85 & 24.94 & 1.4 & 51.76\\
GRB050502B & 5.2 & 50 & 17.72 & 4.72 & 1.58 & 51.77\\
\enddata
\tablecomments{The values of duration $T_{90}$, fluence $S_{\gamma}$ (15-150~keV), and
photon index $\Gamma$ are taken from the third Swift BAT GRB catalogue \citep{2016ApJ...829....7L}
if available, otherwise from the Swift GRB table.
The luminosity $L_{\mathrm{iso}}$ are estimated in the 45-450~keV energy range.
References for the redshift are given as follows:
(1) \citet{2025GCN.42855....1A}; (2) \citet{2025GCN.42350....1I}; (3) \citet{2025GCN.41997....1C}; (4) \citet{2025GCN.41160....1T};
(5) \citet{2025GCN.38934....1M}; (6) \citet{2024GCN.37866....1X}; (7) \citet{2025AA...703A..49B}; (8) \citet{2025arXiv250604340S};
(9) \citet{2023GCN.33749....1A}; (10) \citet{2023GCN.33629....1A}; (11) \citet{2023GCN.33187....1M}; (12) \citet{2022GCN.32928....1I};
(13) \citet{2022GCN.32079....1F}; (14) \citet{2022GCN.31480....1P}; (15) \citet{2022GCN.31359....1F}; (16) \citet{2023AA...671A..84S};
(17) \citet{2021GCN.30164....1Z}; (18) \citet{2020GCN.29100....1M}; (19) \citet{2020GCN.28650....1D}; (20) \citet{2019GCN.25956....1D};
(21) \citet{2019MNRAS.483.5380T}; (22) \citet{2019AA...623A..92S}; (23) \citet{2016GCN.19245....1D}; (24) \citet{2018AA...609A..62B};
(25) \citet{2015GCN.18598....1B}; (26) \citet{2015GCN.17336....1W}; (27) \citet{2014GCN.16301....1C}; (28) \citet{2015AA...581A..86M};
(29) \citet{2015ApJ...804...51C}; (30) \citet{2014GCN.15936....1J}; (31) \citet{2013ApJ...774...26C}; (32) \citet{2013GCN.14634....1S};
(33) \citet{2018ApJ...865..107T}; (34) \citet{2012GCN.13562....1T}; (35) \citet{2014ApJ...781....1L}; (36) \citet{2013MNRAS.428.3590T};
(37) \citet{2012AA...548A..11D}; (38) \citet{2011ApJ...736....7C}; (39) \citet{2009Natur.461.1254T}; (40) \citet{2010AA...522A..20D};
(41) \citet{2010AA...512L...3P}; (42) \citet{2009ApJS..185..526F}; (43) \citet{2009AA...497..729F}; (44) \citet{2007ApJ...663L..57P};
(45) \citet{2006GCN..4815....1B}; (46) \citet{2016ApJ...817....7P}; (47) \citet{2006Natur.440..184K}; (48) \citet{2006AA...447..897J};
(49) \citet{2006ApJ...642..979B}; (50) \citet{2011AA...526A.154A}.}

\end{deluxetable*}

\end{document}